\documentclass[preprint]{elsarticle}

\usepackage[utf8]{inputenc}
\usepackage{todonotes}
\usepackage{url}
\usepackage{amsmath}
\usepackage{amssymb}
\usepackage{xspace}
\usepackage{xr}
\usepackage[colorlinks=true,allcolors=blue]{hyperref}
\usepackage[capitalize]{cleveref}
\usepackage{siunitx}
\usepackage{subcaption}
\usepackage[section]{placeins}
\usepackage{lineno}
\usepackage{mathtools}
\usepackage[toc,page]{appendix}
\usepackage{amsthm}

\newtheorem{remark}{Remark}
\newtheorem{result}{Result}
\newtheorem*{propos}{Proposition}

\numberwithin{equation}{section}

\graphicspath{{./img/}}

\newcommand{\dumux}{DuMu\textsuperscript{x}\xspace}

\renewcommand{\div}{\nabla\!\cdot}
\newcommand{\grad}{\nabla}

\renewcommand{\u}{u}
\newcommand{\avg}[2]{\Pi_{#2}{#1}}
\newcommand{\per}{P}
\renewcommand{\vec}[1]{\boldsymbol{#1}}

\newcommand*\patchAmsMathEnvironmentForLineno[1]{%
	\expandafter\let\csname old#1\expandafter\endcsname\csname #1\endcsname
	\expandafter\let\csname oldend#1\expandafter\endcsname\csname end#1\endcsname
	\renewenvironment{#1}%
	{\linenomath\csname old#1\endcsname}%
	{\csname oldend#1\endcsname\endlinenomath}}%
\newcommand*\patchBothAmsMathEnvironmentsForLineno[1]{%
	\patchAmsMathEnvironmentForLineno{#1}%
	\patchAmsMathEnvironmentForLineno{#1*}}%
\AtBeginDocument{%
	\patchBothAmsMathEnvironmentsForLineno{equation}%
	\patchBothAmsMathEnvironmentsForLineno{align}%
	\patchBothAmsMathEnvironmentsForLineno{flalign}%
	\patchBothAmsMathEnvironmentsForLineno{alignat}%
	\patchBothAmsMathEnvironmentsForLineno{gather}%
	\patchBothAmsMathEnvironmentsForLineno{multline}%
}

\begin{document}

\title{Nonlinear mixed-dimension model for embedded tubular networks
with application to root water uptake}
\author[1,2]{Timo Koch\corref{cor1}}
\ead{timokoch@math.uio.no}
\author[2]{Hanchuan Wu}
\author[2]{Martin Schneider}

\cortext[cor1]{Corresponding author}
\address[1]{Department of Mathematics, University of Oslo, Postboks 1053 Blindern, 0316 Oslo, Norway}
\address[2]{Department of Hydromechanics and Modelling of Hydrosystems, University of Stuttgart, Pfaffenwaldring 61, 70569 Stuttgart, Germany}

\begin{abstract}
We present a numerical scheme for the solution of nonlinear mixed-dimensional PDEs
describing coupled processes in embedded tubular network system in exchange with a bulk domain.
Such problems arise in various biological and technical applications such as in the modeling of root-water uptake,
heat exchangers, or geothermal wells. The nonlinearity appears in form of
solution-dependent parameters such as pressure-dependent permeability or temperature-dependent
thermal conductivity. We derive and analyse a numerical scheme based on distributing
the bulk-network coupling source term by a smoothing kernel with local support.
By the use of local analytical solutions, interface unknowns and fluxes
at the bulk-network interface can be accurately reconstructed from
coarsely resolved numerical solutions in the bulk domain.
Numerical examples give confidence in the robustness of the method and
show the results in comparison to previously published methods. The new method
outperforms these existing methods in accuracy and efficiency. In a
root water uptake scenario, we accurately
estimate the transpiration rate using only
a few thousand 3D mesh cells and a structured cube grid
whereas other state-of-the-art numerical schemes
require millions of cells and local grid refinement to reach comparable accuracy.

\end{abstract}


\maketitle

\section{Introduction}

\begin{figure}
	\centering
	\includegraphics[width=0.6\textwidth]{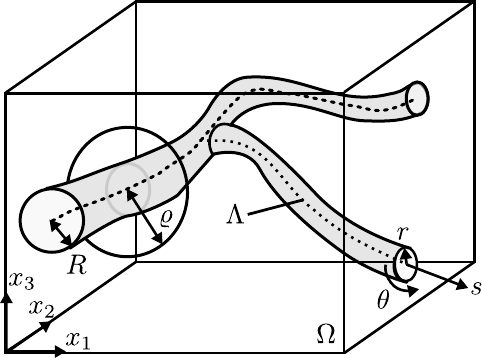}
	\caption{\textbf{Embedded tubular network system.} The network of line segments $\Lambda$ with
	radius function $R$ is embedded into the bulk domain $\Omega$. Both domains are equipped
	with (local) coordinate systems. Source terms coupling equations on $\Lambda$ and $\Omega$
	are distributed in the neighborhood of $\Lambda$ in $\Omega$, i.e. $r \leq \varrho(s)$. This technique allows to bridge
	the dimensional gap between the one-dimensional equations describing processes in $\Lambda$ and
	the three-dimensional equations describing processes in $\Omega$.}
	\label{fig:network}
\end{figure}

Nonlinear elliptic equations arise in the description of fluid flow in porous media where permeability depends
on the water pressure (e.g. Richards' equation) or the description of heat conduction where the thermal conductivity depends
on temperature. In this work, we discuss a numerical scheme to solve such equations in the presence
of an embedded thin tubular transport system exchanging mass or energy with the embedding bulk domain.
This exchange is modeled by local source terms and results in coupled systems of mixed-dimensional
partial differential equations. A motivating example is the simulation of root water uptake from soil
when considering complex three-dimensional root network architectures explicitly~\citep{Dunbabin2013},
which allows to predict the complex water distribution in the soil around roots.

Consider a domain $\Omega \subset \mathbb{R}^3$ with an embedded tubular network.
The centerline is a network
of curves connected at branching points and is denoted by $\Lambda$, as shown in~\cref{fig:network}.
In this work, we want to discuss
the stationary mixed-dimensional nonlinear equation systems of the form
\begin{align}
	\label{eq:nonlineardiffusion-ps}
	-\div \left( D_b(\hat{u}_b) \grad \hat{u}_b \right) &=  q\delta_\Lambda &\text{in} \quad \Omega, \\
	\label{eq:1d-ps}
	- {\partial_s}\left( D_e {\partial_s} u_e \right) &= -q & \text{on} \quad \Lambda, \\
	\label{eq:source-ps}
	q &= -|\per| \gamma \left( \avg{\hat{u}_b}{\per} - u_e \right),
\end{align}
where $s$ is a local coordinate on $\Lambda$ that has a unique mapping to some position $\vec{x} \in \mathbb{R}^3$.
The operators $\div$ and $\grad$ are the spatial divergence and gradient operators in $\Omega$,
and $\partial_s$ denotes the derivative in direction of $s$.
The scalar unknowns in $\Omega$ and $\Lambda$ are denoted
by $\hat{u}_b$ ($b$ for ``bulk'') and $u_e$ ($e$ for ``embeddded'').
The diffusion coefficients $D_b > 0$ and $D_e > 0$ are positive and continuous and $D_b$ is a (possibly nonlinear)
function of $u_b$. In the source term, $\gamma$ is a diffusive permeability
and $\per = \per(s)$ is the cross-sectional tube perimeter at $s$.
Moreover, $\avg{\bullet}{\per}$ is an average operator such that
$\avg{\hat{u}_b}{\per}$ denotes the average of $\hat{u}_b$ on the perimeter $P$ for a given $s$,
\begin{equation}
\label{eq:avg_operator}
\avg{\bullet}{\per} := \frac{1}{|\per|}\int_{\per}\! \bullet \,\text{d}\zeta,
\end{equation}
where $\zeta$ is some suitable parameterization of $\per$ in $\mathbb{R}^3$.
Furthermore, we introduce
\begin{equation}
	\label{eq:circle_average}
	\hat{u}_b^\bigcirc (s) := \avg{\hat{u}_b}{\per} (s),
\end{equation}
explicitly referring to the average of $\hat{u}_b$.
The delta distribution $\delta_\Lambda$ in \cref{eq:nonlineardiffusion-ps} restricts the source term on the centerline.
It has dimension $L^{-2}$ and the property
\begin{equation}
\int_\Omega q \delta_\Lambda \text{d}x = \int_\Lambda q \text{d}s.
\end{equation}

This mixed-dimensional model based on line sources, \cref{eq:nonlineardiffusion-ps,eq:1d-ps,eq:source-ps},
can be derived from a corresponding model with the
three-dimensional tubular network structure cut out from $\Omega$, by assuming that
the tube radii $R$ are much smaller than average distance between tubes~\citep{d2007multiscale}.

The water distribution around a three-dimensional
root network taking up water from the embedding soil can be modeled by~\cref{eq:nonlineardiffusion-ps}~\citep{Gardner1960,Doussan1998,Javaux2008,Dunbabin2013}.
In this case, the unknowns are hydraulic pressures in root and soil, and $D_b$ corresponds to the hydraulic conductivity.
Soil can be viewed as a three-phasic porous medium composed of the solid matrix and two fluid phases, air and water.
With decreasing water content the soil's hydraulic conductivity decreases drastically, and at the same
time, capillary forces increase which attract water to the hydrophilic solid matrix~\citep{helmig1997multiphase}.
Under the assumptions of local mechanical equilibrium a direct and nonlinear relationship between the local
water pressure and the hydraulic conductivity can be found~\citep{mualem1976,van1980closed}. At low water saturation,
high pressure gradients are necessary to move water in dry soil. According to the cohesion-tension theory~\citep{Tyree1997,Steudle2001},
transpiration at the plant leaves causes a high suction potential in plant roots and plants
can maintain water uptake even in relatively dry soils. The root water uptake rate ($q$)
is proportional to the root-soil pressure difference~\citep{Doussan1998}.
This can cause large local pressure gradients around roots~\citep{Steudle2001,Mai2019} which are difficult
to approximate with standard numerical schemes. We will use root water uptake as the motivational
application and in the numerical examples in this work.

\Cref{eq:nonlineardiffusion-ps,eq:1d-ps,eq:source-ps} also arises for heat conduction problems, for example when modeling
the temperature distribution around geothermal well systems~\citep{McDermott2006,Blcher2010}.
Here, $D_b$ corresponds to the thermal conductivity,
which generally depends on temperature~\citep{Mottaghy2007}.
When modeling the flow field around a well in a confined aquifer under high injection
rates the surrounding rock undergoes deformations.
The hydraulic conductivity of the rock depends on the pore pressure.

The paper is structured as follows. Motivated by the results of \citep{Koch2019a},
in \cref{sec:distributedsource} we propose a distributed source model to replace
\cref{eq:nonlineardiffusion-ps,eq:1d-ps,eq:source-ps}. In \cref{sec:reconstr-model},
we design a numerical scheme to accurately approximate
the interface unknowns and hence \cref{eq:source-ps},
locally for each tubular segment.
The method is based on local analytical solution obtained by means of Kirchhoff transformation.
In \cref{sec:numerical}, we show and discuss numerical results.
In particular, in \cref{sec:convergence,sec:convergence_parallel}
we investigate the reconstruction scheme numerically for a series of
carefully constructed verification scenarios.
Finally, in \cref{sec:rootnetwork}, we simulate root water uptake
with a realistic root network obtained from MRI measurement
and compare our results against a numerical reference solution obtained with
state-of-the-art methods.

\section{The distributed source model}
\label{sec:distributedsource}

The introduced model formulation, \cref{eq:nonlineardiffusion-ps,eq:1d-ps,eq:source-ps},
leads to solutions $\hat{u}_b$ which exhibit singularities
on $\Lambda$. It is therefore difficult to construct efficient and accurate numerical schemes for solving
\cref{eq:nonlineardiffusion-ps,eq:1d-ps,eq:source-ps}~\citep{koeppl2018,Laurino2019,Koch2019a}.
However, for a precise description of the source term coupling the network and bulk domain,
it is crucial to accurately approximate the solution in the neighborhood of the network.
Koch et al.~\citep{Koch2019a} suggest to solve a modified problem
\begin{align}
  \label{eq:nonlineardiffusion}
  -\div \left( D_b(u_b) \grad u_b \right) &= q\Phi_\Lambda & \text{in} \quad \Omega, \\
  \label{eq:1d}
  - {\partial_s}\left( D_e {\partial_s} u_e \right) &= -q & \text{on} \quad \Lambda, \\
  \label{eq:source}
  q &= -|\per| \gamma ( \hat{u}_b^\bigcirc - u_e ),
\end{align}
where $\Phi_\Lambda$ denotes a set of non-negative kernel functions $\Phi_{\Lambda,i}$
that distribute $q$ around a vessel segment $i$ over a small radially-symmetric tubular support region
with radius $\varrho(s)$, where $\Phi_{\Lambda,i} = 0$ outside the support region (compact support), cf. \cref{fig:network}.

\begin{figure}
	\includegraphics[width=\textwidth]{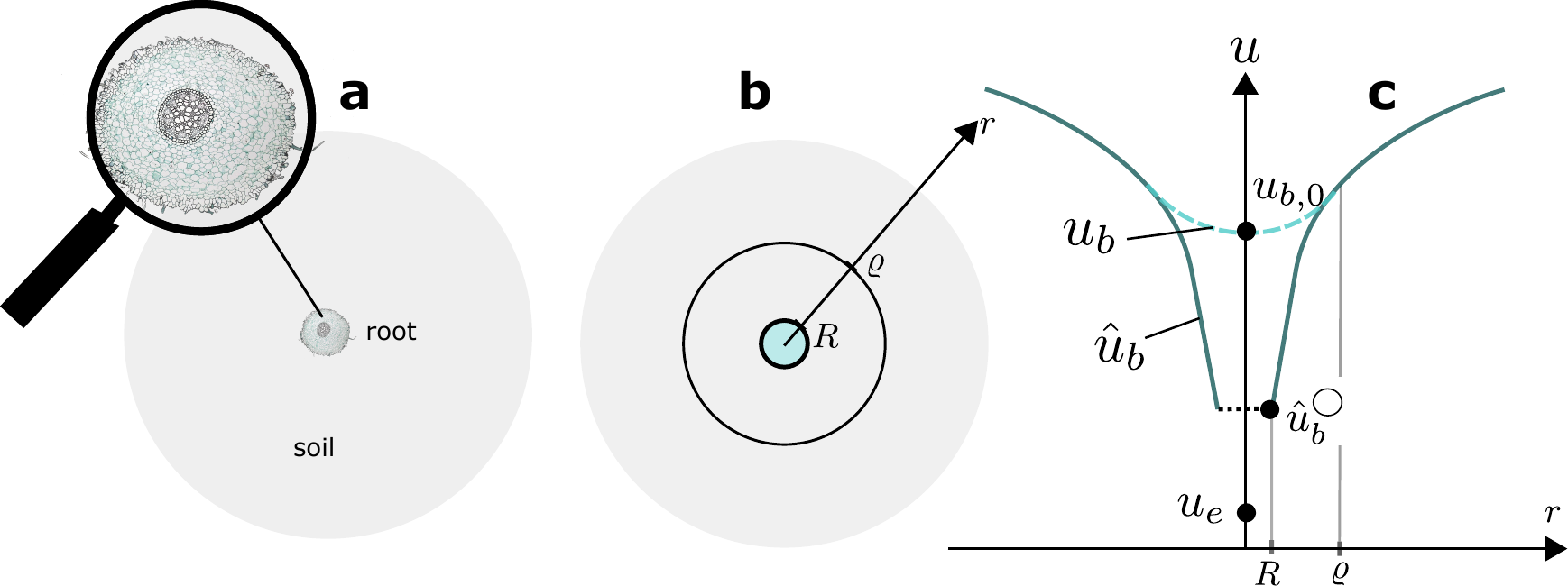}
	 \caption{\textbf{The distributed source model.} (A) Exemplary application: root water uptake from soil. Cross-sectional cut through root embedded in soil.
	 (B) Terminology for radial symmetric setup, root radius and kernel support radius for local source distribution.
	 (C) Effect of the regularization kernel on the local soil pressure solution.
	 Symbols: $\u_{b,0}$, regularized unknown at the tube centerline;
	 $\hat{u}_b^\bigcirc$, (average) bulk interface quantity (e.g. soil pressure);
	 $u_e$, physical quantity inside the tube (e.g. root pressure);
	 $R$, tube radius; $\varrho$, kernel support radius.}
	 \label{fig:kernelscheme}
\end{figure}

\begin{remark}
We choose kernel functions $\Phi_{\Lambda,i}(s)$ along each segment $i$ in the form~\cite{Koch2019a}
\begin{equation}
\Phi_{i}(\varrho) = \varrho^{-2} \varphi(r\varrho^{-1}) \quad \text{with}
\quad  \int\displaylimits_0^{2\pi}\!\int\displaylimits_0^{\varrho(s)} \Phi_{\Lambda,i} r\;\text{d}r\text{d}\theta = 1 \quad \forall s,
\end{equation}
where the function $\varphi$ is a positive symmetric mollifier~\citep{Friedrichs1944,Friedrichs1953}
defined on a disc perpendicular to the vessel segment, cf. \cref{fig:network}.
An example for such kernel functions is given in~\cref{eq:constkerneldefinition}.
The kernel functions $\Phi_\Lambda$ have dimension $L^{-2}$ ($L$: length) and bridge the dimensional gap of $2$ between
\cref{eq:nonlineardiffusion} and \cref{eq:1d}.
In this work we use a uniform distribution,
\begin{equation}
\label{eq:constkerneldefinition}
	\Phi_{\Lambda,i}(r) = \begin{cases}
		\frac{1}{\pi\varrho^2} & r \leq \varrho,\\
		0 & r > \varrho. \end{cases}
\end{equation}
\end{remark}
\begin{remark}
In the limit $\varrho \rightarrow 0$, \cref{eq:nonlineardiffusion,eq:1d,eq:source}
reduce to \cref{eq:nonlineardiffusion-ps,eq:1d-ps,eq:source-ps}. For $\varrho \leq R$,
$\hat{u}_b^\bigcirc$ can be approximated by $\avg{u_b}{\per}$. However, for $\varrho > R$,
this approximation is poor as the distribution kernel leads to a locally regularized solution $u_b \neq \hat{u}_b$
(visualized in \cref{fig:kernelscheme}).
\end{remark}

When simulating systems with large networks of thousands of tubes (e.g. root systems)
resolving the local solution around each network segment requires fine local computational meshes.
The kernel distributes the source or sink term in a local neighborhood
around network segments. This leads to a smooth $u_b$
which is easy to approximate by standard numerical schemes, see \cref{fig:kernelscheme}
for schematic representation at the example of a root segment cross-section.
Unfortunately, the value of $u_b$ at the tube-bulk interface does not correspond to
the interface value of the line source model \cref{eq:nonlineardiffusion-ps,eq:1d-ps,eq:source-ps} anymore, i.e. $\avg{u_b}{\per} \neq \hat{u}_b^\bigcirc$.
In the following, we exploit the fact that we know \emph{how} the
introduced kernel function modifies the local solution around a single isolated tube segment.
The design of a method to reconstruct $\hat{u}_b^\bigcirc$ accurately from the smooth solution $u_b$---
in the presence of the nonlinearity introduced by the diffusion coefficient $D_b$---
is the main contribution of this work.

Finally, we note that the linear case, i.e. $D_b = \text{const.}$ is discussed in \citep{Koch2019a}.
For the particular case of root water uptake, local corrections based on the analytical
solutions of the Richards equation have been proposed in \citep{Schroeder2008localsoil}.
However, the scheme is only presented in the discrete setting for a single voxel.
The authors of \citep{Mai2019} suggest to solve local radial-symmetric problems at every root segment, introducing
additional unknowns. Again, the method is only presented in the discrete setting.
In this work, we follow \citep{Koch2019a} and present a model formulated on the
continuous setting which allows generalization to any suitable discretization method
and simplifies the analysis of possible sources of errors.
Moreover, the source distribution kernel allows to control the
accuracy of the interface reconstruction in interplay with the mesh size.
Apart from getting rid of singularities, the problem formulation \labelcref{eq:nonlineardiffusion,eq:1d,eq:source}
has the advantage that it does not require that the computational grid resolves the
length scale $R$ to yield accurate approximations of $q$~\citep{koeppl2018}, but
relaxes this requirement to the grid being required to resolve the length scale $\varrho$,
a selectable model parameter. It has been shown~\citep{Koch2019a} that this formulation allows
to significantly reduce the error in $q$, $u_e$, and $u_b$ for coarse grids.

\section{Local reconstruction of interface unknown and flux}
\label{sec:reconstr-model}

In this section we describe a method to accurately reconstruct the interface
unknown $\hat{u}_b^\bigcirc$ for a given tube segment from the evaluation of
the smoothed solution $u_b$ on the centerline $\Lambda$.
The Kirchhoff transformation, well-known from the solution of
heat conduction problems~\citep[][Eq. (10)]{carslaw1992conduction} allows us
to transform \cref{eq:nonlineardiffusion} such that the nonlinearity only
appears in the source term. We then derive a local analytical solution
for $u_b$ depending on tube and kernel radius. From this analytical solution,
we deduct a nonlinear equation to compute $\hat{u}_b^\bigcirc$
from a point evaluation of $u_b$. We conclude by discussing the validity
of the approach for the case of tubular networks.

\subsection{Kirchhoff transformation}
\label{sec:kirchhoff}

\Cref{eq:nonlineardiffusion,eq:1d,eq:source} is a nonlinear equation system, if
the diffusion coefficient $D_b$ depends on $u_b$. Let us
introduce the following Kirchhoff transformation~\citep{Berninger2011}
\begin{equation}
\label{eq:kirchhoff}
T: \u_b \mapsto \psi = \int_0^{\u_b}\! D_b(\tilde{\u_b}) \,\text{d}\tilde{\u_b},
\end{equation}
and we define $\hat{\psi}$ analogously in terms of $\hat{u}_b$.
The chain rule yields
\begin{equation}
	\label{eq:kirchhoff_deriv}
	\grad \psi = D_b(\u_b) \grad u_b,
\end{equation}
which allows us to rewrite the left-hand-side of \cref{eq:nonlineardiffusion},
\begin{equation}
\label{eq:poisson_psi}
-\nabla\cdot \nabla \psi = q\Phi_\Lambda,
\end{equation}
in terms of the transformed variable $\psi$. Note that the left-hand-side is now a linear operator,
and the nonlinearity is contained in the source term
\begin{equation}
\label{eq:transformed source}
	q = -|\per| \gamma (\avg{T^{-1}(\hat{\psi})}{P} - \u_e),
\end{equation}
in terms of the inverse Kirchhoff transformation. This is an essential step,
since \cref{eq:poisson_psi} can be solved analytically
in a simple radially-symmetric setting, cf.~\cref{sec:reconstruction}.

\begin{remark}
Since $D_b$ is assumed positive and continuous, $T$ is a strictly monotonically
increasing function and we can uniquely define its inverse, $T^{-1}$.
However, depending on the choice of $D_b(\u_b)$,
the image of $T(\mathbb{R})$ can be a strict subset ($\psi_c$, $\infty$) of $\mathbb{R}$,
and consequently $T^{-1}$ may be only defined on ($\psi_c$, $\infty$)~\citep{Berninger2011}.
This means not all solutions $\psi$ of \cref{eq:poisson_psi} have corresponding solutions $u_b$.
\end{remark}

\begin{figure}[htb!]
	\centering
	\includegraphics[width=1.0\textwidth]{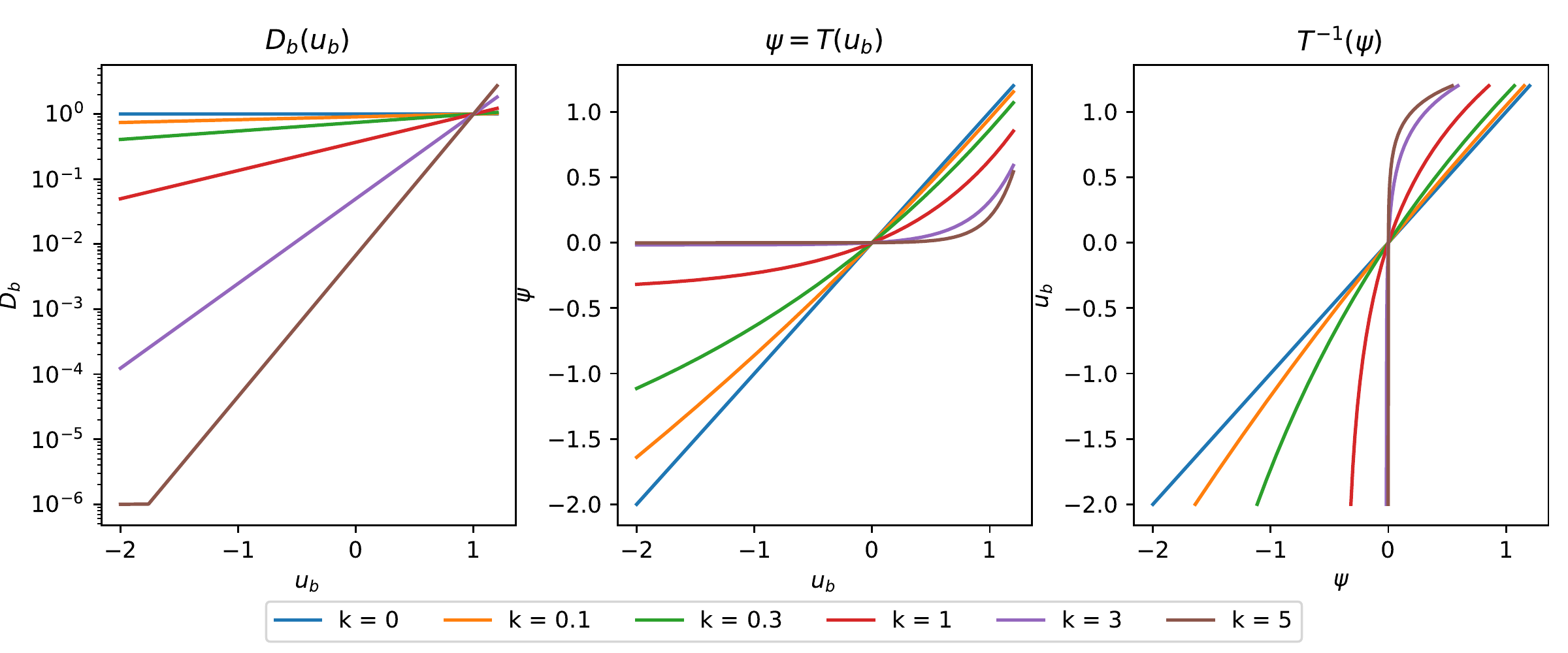}
	\caption{\textbf{An exponential diffusion coefficient function.} Used for model verification tests in this work.
	The diffusion coefficient is given by $D_b(u_b) = \max(D_0 \exp\{ k(u_b - 1) \}, D_\text{min})$,
	where $D_0 = 1$, and $D_\text{min} = \SI{1e-6}{}$. The middle and right plot show the Kirchhoff transformation,
	$T(u_b)$, as defined in \cref{eq:kirchhoff} and its inverse function, $T^{-1}(\psi)$.
	The analytical expressions for $T(u_b)$ and $T^{-1}(\psi)$ are given in \cref{sec:app:exp}.}
	\label{fig:Dbr}
\end{figure}

For example, for the exponential function $D_b = D_0\exp\{ k (u_b-1) \}$ and $k \in \mathbb{R}^+$,
a function that fulfills our preconditions on $D_b$
and that we will use subsequently in numerical verification tests, this is indeed the case.
There, $\psi_c$ corresponds to $u_b = -\infty$ and is well-defined.
Nevertheless, $T^{-1}$ is typically ill-conditioned around $\psi_c$ (unbounded derivatives).
This is evident in \cref{fig:Dbr} for $k=5$ where it becomes clear that the derivative can get arbitrarily large.
(The same holds true for the
Van Genuchten-Mualem model commonly used for the permeability
in root-soil interaction, see \cref{fig:krS_krPw} at $\theta_r\theta_s^{-1}$ in \cref{sec:rootnetwork}.)

\begin{propos}
Fortunately, this singularity disappears in the non-degenerate case~\citep{Berninger2011}
\begin{equation}
	D_b(u_b) \geq D_\text{min} \quad \text{for some} \quad D_\text{min} > 0.
\end{equation}
In this case both $T$ and $T^{-1}$ are defined on all of $\mathbb{R}$.
For instance, this property can be achieved for the exponential function $D_b$ by using
$D_\text{min}$ as its lower bound, i.e. $D_b := \max( D_0\exp\{ k (u_b-1) \}, D_\text{min})$.
\end{propos}
The functions shown in \cref{fig:Dbr} are regularized in this way using $D_\text{min} = \SI{1e-6}{}$ (see $k=5$).

\subsection{Local cylinder model and interface reconstruction}
\label{sec:reconstruction}

Firstly, let us consider an infinitely long straight cylinder of radius $R$
embedded in an infinite domain $\Omega_\infty \subset \mathbb{R}^3$ with $u_e$ being a given constant.
Solving \cref{eq:nonlineardiffusion,eq:1d,eq:source} then reduces to finding radially symmetric solutions by solving
\begin{equation}
	\label{eq:polarcoodinatenonlineardiffusion}
	-\frac{1}{r}\frac{\partial}{\partial r}\left(rD_b(u_b)\frac{\partial u_b}{\partial r}\right) = q(u_e, \hat{u}_b^\bigcirc) \Phi_\Lambda,
\end{equation}
on a cross-sectional plane in local cylinder coordinates.
By using the Kirchhoff transformation, \cref{eq:kirchhoff}, and the definition of $\hat{u}_b^\bigcirc$, \cref{eq:circle_average}, we obtain
\begin{equation}
	\label{eq:tpolarcoodinatenonlineardiffusion_trans}
	-\frac{1}{r}\frac{\partial}{\partial r}\left(r\frac{\partial \psi}{\partial r}\right) = q(u_e, \avg{T^{-1}(\hat{\psi})}{\per}) \Phi_\Lambda.
\end{equation}
As discussed above, $q$ is now a nonlinear function of $\hat{\psi}$
(which solves \cref{eq:tpolarcoodinatenonlineardiffusion_trans} for $\varrho \rightarrow 0$).
Using the uniform distribution kernel, \cref{eq:constkerneldefinition}, which is only non-zero
in the local support region defined by the kernel radius $\varrho \geq R$,
we obtain an analytical solution,
\begin{equation}
\label{eq:analytical_radsym}
\psi(r) = \begin{cases}  \hat{\psi}^\bigcirc - \frac{q}{2\pi }\left[ \frac{r^2}{2\varrho^2} + \ln\left(\frac{\varrho}{R}\right) - \frac{1}{2} \right] & r \leq \varrho \\
\hat{\psi}^\bigcirc - \frac{q}{2\pi}\left[ \ln\left(\frac{r}{R}\right) \right] & r > \varrho,
\end{cases}
\end{equation}
in terms of the transformed variable $\psi$ and
\begin{equation}
\hat{\psi}^\bigcirc := \avg{\hat{\psi}}{P} \overset{\varrho \leq R}{=} \avg{\psi}{P}.
\end{equation}

Continuing in this setting and evaluating \cref{eq:analytical_radsym} at $r=\delta$,
where $0 \leq \delta < \varrho$ denotes a distance to some point close to the tube centerline ($r=0$),
yields
\begin{equation}
\label{eq:psibar}
\hat{\psi}^\bigcirc = \psi_\delta + \frac{q}{2\pi}\left[ \frac{\delta^2}{2\varrho^2} + \ln\left(\frac{\varrho}{R}\right) - \frac{1}{2} \right],
\end{equation}
where we introduced the symbol $\psi_\delta := \psi(r=\delta)$. We note that
we evaluated $\psi$ in the region regularized by the source distribution kernel.
Moreover, recall that the coupling source term is given by
\begin{equation}
\label{eq:source2d}
q =-|\per| \gamma (\hat{u}_b^\bigcirc - \u_e),
\end{equation}
and with the Kirchhoff transformation and radial symmetry, we know that
\begin{equation}
\label{eq:transforms}
\psi_\delta = T(\u_{b,\delta}) \quad \text{and} \quad \hat{\psi}^\bigcirc = \avg{\hat{\psi}}{P} = T(\avg{\hat{u}_b}{P}) = T(\hat{u}_b^\bigcirc).
\end{equation}

\begin{result}
If we can estimate $\u_{b,\delta}$
(for example by the discrete cell value in a finite volume discretization),
we can find the interface pressure $\hat{u}_b^\bigcirc$ by solving
a nonlinear equation composed of \cref{eq:psibar,eq:source2d,eq:transforms},
\begin{equation}
  \label{eq:nleqifpressure}
T(\u_{b,\delta}) - T(\hat{u}_b^\bigcirc) - \frac{|P|\gamma}{2\pi} \left[ \frac{\delta^2}{2\varrho^2} + \ln\left(\frac{\varrho}{R}\right) - \frac{1}{2} \right] \left(\hat{u}_b^\bigcirc - \u_e\right) = 0,
\end{equation}
which then allows to compute the source term $q$, given by \cref{eq:source2d}.
This means that an accurate approximation of $\u_{b,\delta}$ and $u_e$ is sufficient to compute
a good approximation of $q(\hat{u}_b^\bigcirc, u_e)$.
\end{result}
This also suggests that we can indeed solve \cref{eq:nonlineardiffusion,eq:1d,eq:source} instead of
\cref{eq:nonlineardiffusion-ps,eq:1d-ps,eq:source-ps} while retaining a good approximation
of $q$, $u_e$ and $\hat{u}_b$ (for $r > \varrho$).

\begin{remark}
In the case that $D_b$ is a constant, \cref{eq:nleqifpressure}
can be explicitly solved for $\hat{u}_b^\bigcirc$ as shown in \citep{Koch2019a}.
\end{remark}
\begin{remark}
To see that \cref{eq:nleqifpressure} has a unique solution,
it is sufficient to show that the expression on the left side is monotone with respect to $\hat{u}_b^\bigcirc$. In general, this only holds under some conditions.
First, we assume that the kernel radius is chosen large enough such that $\ln\left(\frac{\varrho}{R}\right) \geq 0.5$.
With this assumption, the third term is a monotonically decreasing (linear) function of $\hat{u}_b^\bigcirc$.
Due to the assumption of positive diffusion coefficients, the functional $-T$ is also monotonically decreasing.
It follows the monotonicity of the entire left hand side expression in \cref{eq:nleqifpressure}
(constant terms do not influence the monotonicity), and \cref{eq:nleqifpressure} therefore has a unique solution.
\end{remark}

For the general three-dimensional case,
the second equality in \cref{eq:transforms} is only an approximation.
The equality holds here due to the radial symmetry of the solution.
This condition is violated in the presence of multiple arbitrarily-oriented
tubes as they readily occur in tubular network structures (microvasculature, root systems, fibre networks).
We motivate in the next section why this error is expected to be small.

\subsection{Multiple interacting parallel tubes}
\label{sec:paralleltubes}
Let us consider the case of many parallel tubes where the solution $\u_b$
only varies linearly along the tubes. Hence, we can consider a two-dimensional
cross-sectional plane and denote with $\vec{x} \in \mathbb{R}^2$
a position on this plane. Due to the linearity of the Laplace operator in
\cref{eq:poisson_psi}, we can obtain a general solution for $\psi$ by
superposition
\begin{equation}
\psi = \sum_{j=1}^N \psi_j + H, \quad \psi_j = - q_j f_j(\vec{x}),
\label{eq:psisuperpos}
\end{equation}
where $H$ is some
harmonic function (for example chosen such that some boundary conditions are satisfied),
\begin{align}
\label{eq:paralleltubes_sources}
q_j &= -|\per_j| \gamma_j (\hat{u}_{b,j}^\bigcirc - \u_{e} (\vec{x}_j)),
\\
\label{eq:fjSol}
f_j(\vec{x}) &= \begin{cases} \frac{1}{2\pi}\left[ \frac{|| \vec{x} - \vec{x}_j ||_2^2}{2\varrho_j^2} + \ln\left(\frac{\varrho_j}{R_j}\right) - \frac{1}{2} \right] & ||\vec{x} - \vec{x}_j||_2 \leq \varrho_j, \\
\frac{1}{2\pi }\left[ \ln\left(\frac{|| \vec{x} - \vec{x}_j ||_2}{R_j}\right) \right] & \text{else.}
\end{cases}
\end{align}
We remark that $\hat{u}_{b,j}^\bigcirc$ in $q_j$ depends on
contributions from all partial solutions $\psi_j$.
%
%
To simplify notation, we assume $\varrho_i \leq R_i$ in the following ($\hat{u}_{b,i}^\bigcirc = \avg{\hat{u}_b}{\per_i} = \avg{\u_b}{\per_i}$).
Applying the linear averaging operator (defined in \cref{eq:avg_operator})
on both sides of \cref{eq:psisuperpos} and assuming that the tubes do not overlap yields
\begin{align}
\label{eq:averaging_superpos}
\avg{\psi}{\per_i} &= \avg{\psi_i}{\per_i} + \sum_{j=1, j \not = i}^N  \psi_j(\vec{x}_i) + H(\vec{x}_i) \nonumber\\
                   &= \avg{\psi_i}{\per_i} + \psi(\vec{x}_i) - \psi_i(\vec{x}_i),
\end{align}
where we used the mean value theorem for
harmonic functions, i.e. $\avg{\psi_j}{\per_i} = \psi_j(\vec{x}_i)$.
From this, it follows that
\begin{align}
\psi(\vec{x}_i) - \avg{\psi}{\per_i} &=  \psi_i(\vec{x}_i) - \avg{\psi_i}{\per_i} \nonumber\\
&= q_i (\avg{f_i}{\per_i} - f_i(\vec{x}_i)) \nonumber\\
&= |P_i| \gamma_i (\hat{u}_{b,i}^\bigcirc - \u_{e} (\vec{x}_i)) (f_i(\vec{x}_i) - \avg{f_i}{\per_i}) \nonumber\\
&= |P_i| \gamma_i (\hat{u}_{b,i}^\bigcirc - \u_{e} (\vec{x}_i)) f_i(\vec{x}_i),\label{eq:nonlinear:mv}
\end{align}
%
since the source term $q_i$ is independent of $\vec{x}_i$ and $\avg{f_i}{\per_i} = 0$.
Finally, assuming that
\begin{equation}
\label{eq:main_assumption_mvt}
\avg{\psi}{\per_i} = \avg{T(\u_b)}{\per_i} \approx T(\avg{\u_b}{\per_i}) = T(\hat{u}_b^\bigcirc),
\end{equation}
yields the reconstruction equation
\begin{equation}
\label{eq:nleqifpressure-mult}
T(\u_b(\vec{x_i})) - T(\tilde{u}_{b,i}^\bigcirc) - |P_i| \gamma_i (\tilde{u}_{b,i}^\bigcirc - \u_e(\vec{x_i})) f_i(\vec{x}_i) = 0,
\end{equation}
which is equivalent to \cref{eq:nleqifpressure} with $\delta = 0$.
The $\tilde{u}_{b_i}^\bigcirc \approx \hat{u}_{b,i}^\bigcirc$
resulting from solving \cref{eq:nleqifpressure-mult} is an approximation due to \cref{eq:main_assumption_mvt}.
This extends our result from the single vessel case to multiple parallel vessels,
but with the introduction of some approximation error due to the nonlinearity of $D_b$.
Since \cref{eq:nleqifpressure-mult} only requires point evaluations of $u_b$,
the result extends to $\varrho_i > R_i$ as long as the kernel support regions do not overlap.

\subsubsection{Error estimate for the approximation of the average operator}
\label{sec:errorestimate}
In the following, we estimate the error associated
with approximation \eqref{eq:main_assumption_mvt}
in the reconstruction of the interface unknown.
Let us assume that $\psi(\vec{x}_i)$ and $\u_{e} (\vec{x}_i)$ are given and denote with $\u_b$ the exact solution.
Again, to simplify notation, we assume $\varrho_i \leq R_i$.
Subtracting \cref{eq:nonlinear:mv} from \cref{eq:nleqifpressure-mult} yields
\begin{align}
\avg{\psi}{\per_i}  - T(\tilde{u}_{b,i}^\bigcirc) = F_i  (\hat{u}_{b,i}^\bigcirc - \tilde{u}_{b,i}^\bigcirc),
\label{eq:nonlinear:mv:diff}
\end{align}
where $F_i:= -|P_i| \gamma_i f_i(\vec{x}_i) \geq 0$.
Assuming that $D_b \in C^1(U)$ (i.e. $T\in C^2(U)$), for some $U \subset \mathbb{R}$ such that $\u_b(\Omega) \subset U$, and using Taylor's Theorem results in
\begin{align}
\psi = T(u) = T(\hat{u}_{b,i}^\bigcirc) + T^\prime (\hat{u}_{b,i}^\bigcirc)  (u - \hat{u}_{b,i}^\bigcirc)
    +  \int\limits^u_{\hat{u}_{b,i}^\bigcirc}  T^{\prime \prime} (\tilde{u}) (u - \tilde{u}) \, \text{d} \tilde{u},
\end{align}
for any $u \in U$. Inserting the exact solution and applying the averaging operator on both sides give
\begin{align}
	\avg{\psi}{\per_i} = T(\hat{u}_{b,i}^\bigcirc) +  \avg{}{\per_i} \left( \int\limits^{\u_b}_{\hat{u}_{b,i}^\bigcirc}  T^{\prime \prime} (\tilde{u}) (\u_b - \tilde{u}) \, \text{d} \tilde{u} \right),
\end{align}
and inserting this expression into \cref{eq:nonlinear:mv:diff} results in
\begin{equation}
F_i  (\hat{u}_{b,i}^\bigcirc  - \tilde{u}_{b,i}^\bigcirc) = T(\hat{u}_{b,i}^\bigcirc) - T(\tilde{u}_{b,i}^\bigcirc)
+  \avg{}{\per_i} \left( \int\limits^{\u_b}_{\hat{u}_{b,i}^\bigcirc}  T^{\prime \prime} (\tilde{u})  (\u_b - \tilde{u}) \, \text{d} \tilde{u} \right).
\end{equation}
The above equation can be equivalently written as
\begin{equation}
\begin{split}
\int\limits_{\hat{u}_{b,i}^\bigcirc}^{\tilde{u}_{b,i}^\bigcirc}  (T^\prime(\tilde{\u_b}) - F_i)\,  \,\text{d}\tilde{\u_b}
&=  \avg{}{\per_i} \left( \int\limits^{\u_b}_{\hat{u}_{b,i}^\bigcirc}  T^{\prime \prime} (\tilde{u})  (\u_b - \tilde{u}) \, \text{d} \tilde{u} \right).
\end{split}
\end{equation}
By assuming that there exists some constant $\tilde{C}_i$, independent of $R_i$, such that
\begin{equation}
\left| \,\int\limits_{\hat{u}_{b,i}^\bigcirc }^{\tilde{u}_{b,i}^\bigcirc}  (T^\prime(\tilde{\u_b}) - F_i)\,  \,\text{d}\tilde{\u_b}  \right| \geq \tilde{C}_i \left| \tilde{u}_{b,i}^\bigcirc - \hat{u}_{b,i}^\bigcirc  \right|,
\end{equation}
which for example holds if
$D_b^{-1}(F_i) \not \in [\min( \tilde{u}_{b,i}^\bigcirc,\hat{u}_{b,i}^\bigcirc), \max(\tilde{u}_{b,i}^\bigcirc,\hat{u}_{b,i}^\bigcirc)] $,
it holds that
\begin{align}
\tilde{C}_i \left| \tilde{u}_{b,i}^\bigcirc -\hat{u}_{b,i}^\bigcirc  \right| &\leq \left|  \avg{}{\per_i} \left( \int\limits^{\u_b}_{\hat{u}_{b,i}^\bigcirc}  T^{\prime \prime} (\tilde{u})  (\u_b - \tilde{u}) \, \text{d} \tilde{u} \right) \right| \\
&\leq 0.5\lVert D^\prime_b \rVert_{L^\infty(u_b(P_i))} \avg{(\u_b - \hat{u}_{b,i}^\bigcirc)^2}{\per_i}  \\
&\leq 0.5\lVert D^\prime_b \rVert_{L^\infty(u_b(P_i))} (\text{ess} \sup_{P_i} \u_b - \text{ess} \inf_{P_i} \u_b )^2.
\end{align}

If the solution $\u_b$ is sufficiently smooth (e.g. $C^1$ or Lipschitz continuous), then we deduce from the inequality above that the error introduced by the approximation $\tilde{u}_{b,i}^\bigcirc$ is $\mathcal{O}(R_i^2)$.
Furthermore, it also shows that there is no error if $D_b = \text{const.}$, i.e. \cref{{eq:nonlineardiffusion}} is a linear diffusion equation, or if $\u_b$ is constant on $P_i$ (i.e. radial symmetric solution).

\begin{remark}
In a numerical scheme, the exact solution $\psi$ is approximated
by some discrete solution $\psi_h$,
for which it holds that $|\psi(\vec{x}_i) - \psi_h(\vec{x}_i)| = \mathcal{O}(h^2)$ ($h$ denotes the discretization length, see \cref{sec:numerical}),
when using a second order scheme.
Therefore, the approximation $\psi_h(\vec{x}_i) \approx \psi(\vec{x}_i)$
introduces an error of $\mathcal{O}(h^2)$,
whereas the error introduced by the approximation
$\avg{\psi}{\per_i} \approx T(\hat{u}_{b,i}^\bigcirc)$
is in the order of $\mathcal{O}(R_{i}^2)$.
This suggests that on grids where $h > R_{i}$,
$\avg{\psi}{\per_i} \approx T(\hat{u}_{b,i}^\bigcirc)$ yields a good approximation,
without being the main source of error. This result is supported by
the numerical results in \cref{sec:numerical}. Since the goal of the kernel method
is to allow coarser grids by choosing kernel support radii
$\varrho > R_{i}$ while maintaining accuracy~\citep{Koch2019a}, grid resolutions with $h > R_{i}$
correspond to the typical use case.
\end{remark}

\subsection{Multiple arbitrarily-oriented tubes}
\label{sec:network}
The numerical method introduced above for single or parallel tubes
can also be applied for the general three-dimensional case with arbitrarily-oriented tubes.
However, for arbitrarily-oriented tubes an additional error is introduced because the mean value property of harmonic
functions, used for $\psi_j$ to derive \cref{eq:nonlinear:mv}, is no longer valid.

The additional error depends on
$|\avg{\psi_j}{\per_i} - \psi_j(\vec{x}_i)| = |q_j| |\avg{f_j}{\per_i} - f_j(\vec{x}_i)|$,
with $j \not = i $ and $f_j$ as defined in \cref{eq:fjSol}.
Assuming that the kernel support regions are
non-overlapping, we apply Taylor's Theorem to deduce the following estimate
\begin{equation}
|\avg{f_j}{\per_i} - f_j(\vec{x}_i)| \leq \frac{C}{2} R_i^2.
\end{equation}
Assuming that a contribution of another tube has the shape of a line source,
the constant $C$ can be computed, yielding
\begin{equation}
|\avg{f_j}{\per_i} - f_j(\vec{x}_i)|
	\leq \frac{R_i^2}{4\pi (\lVert \vec{x}_i - \mathcal{E}_j^\perp(\vec{x}_i)\rVert_2 - R_i)^2 },
\end{equation}
where $B_{R_i}(\vec{x}_i)$ denotes a ball of radius $R_i$ centered at $\vec{x}_i$,
and $\mathcal{E}_j^\perp$ orthogonally projects $\vec{x}$ onto the centerline of tube $j$.
The derivation of the estimate is given in \cref{sec:app:estimate_3d}.
We note that most other tubes in a large network system embedded
in $\Omega$ are far away from the segment $i$, and therefore the error is small. Close
tubes may cause a signification error. However, average distances to the closest neighbor
tube in relevant applications are often $10R_i$ and larger. With these estimates and by using
similar arguments as in \cref{sec:errorestimate},
we conclude that also for the case of arbitrarily-oriented tubes,
the error introduced by reconstructing $\tilde{u}_{b,i}^\bigcirc$
from \cref{eq:nleqifpressure-mult} using approximation \labelcref{eq:main_assumption_mvt}
is not dominant on grids where $h > R_i$.
This conclusion is supported by numerical experiments, e.g.~\cite{Koch2021a}.

Furthermore, the line segments in practical networks are finite and
kinks and bifurcations may introduce additional errors since there
the kernel support regions of two connected vessels may overlap and the assumption of discrete cylinders as segments
may result in errors in the estimation of the network surface area.
In \cite{Koch2021a}, the author proposes
a numerical method that does not suffer from these
errors since the tube interface is explicitly resolved by the three-dimensional
computational mesh. It is shown that the approximation of kinks and bifurcations
with cylinder segments do not introduce significant errors and that these errors
are irrelevant in practical simulations of tissue perfusion or root water uptake.

\section{Numerical results and discussion}
\label{sec:numerical}

The three-dimensional bulk domain $\Omega$ and the network domain $\Lambda$
are spatially decomposed into the meshes $\Omega_h$ and $\Lambda_h$ consisting
of control volumes (cells) $K_\Omega \in \Omega_h$ and $K_\Lambda \in \Lambda_h$, respectively.
The discretization length, i.e. the maximum cell diameter, is denoted as $h$.
To discretize the nonlinear diffusion problem \cref{eq:nonlineardiffusion,eq:1d,eq:source} in space,
a cell-centered finite volume method with a two-point flux approximation is employed \cite{Koch2019a}.
The resulting nonlinear system of equations is solved with Newton's method.
To solve the learnized system of equations within each Newton iteration, we use the same linear solver as described in \cite{Koch2019a},
that is, a stabilized bi-conjugate gradient method with a block-diagonal incomplete LU-factorization.
All presented methods and simulations are implemented using the open-source software framework \dumux~\cite{Dumux32019}
with the network grid implementation \texttt{dune-foamgrid}~\cite{foamgrid} for representing the embedded network domain.

For general nonlinear constitutive models
(e.g. the Van Genuchten curves in \cref{sec:rootnetwork}),
an analytical expression for the Kirchhoff
transformation \cref{eq:poisson_psi} and its inverse cannot be found such that it has to be calculated numerically. Here, we use numerical integration based on the
double exponential transformation~\cite{Mori2001doublexp} and Brent's method
for the inverse transformation.  For fast evaluation of the inverse transformation,
the functional $T^{-1}(\psi)$ is replaced by a lookup table with
local linear interpolation and a high sampling rate.
The nonlinear interface reconstruction, \cref{eq:nleqifpressure} (single tube) or \cref{eq:nleqifpressure-mult} (multiple tubes), is solved
using Brent's method.

\subsection{Analytical solutions for multiple parallel tubes}
\label{sec:analytical_multiple}

\begin{figure}[!htb]
	\includegraphics[width=1.0\textwidth,trim=60 50 50 50,clip]{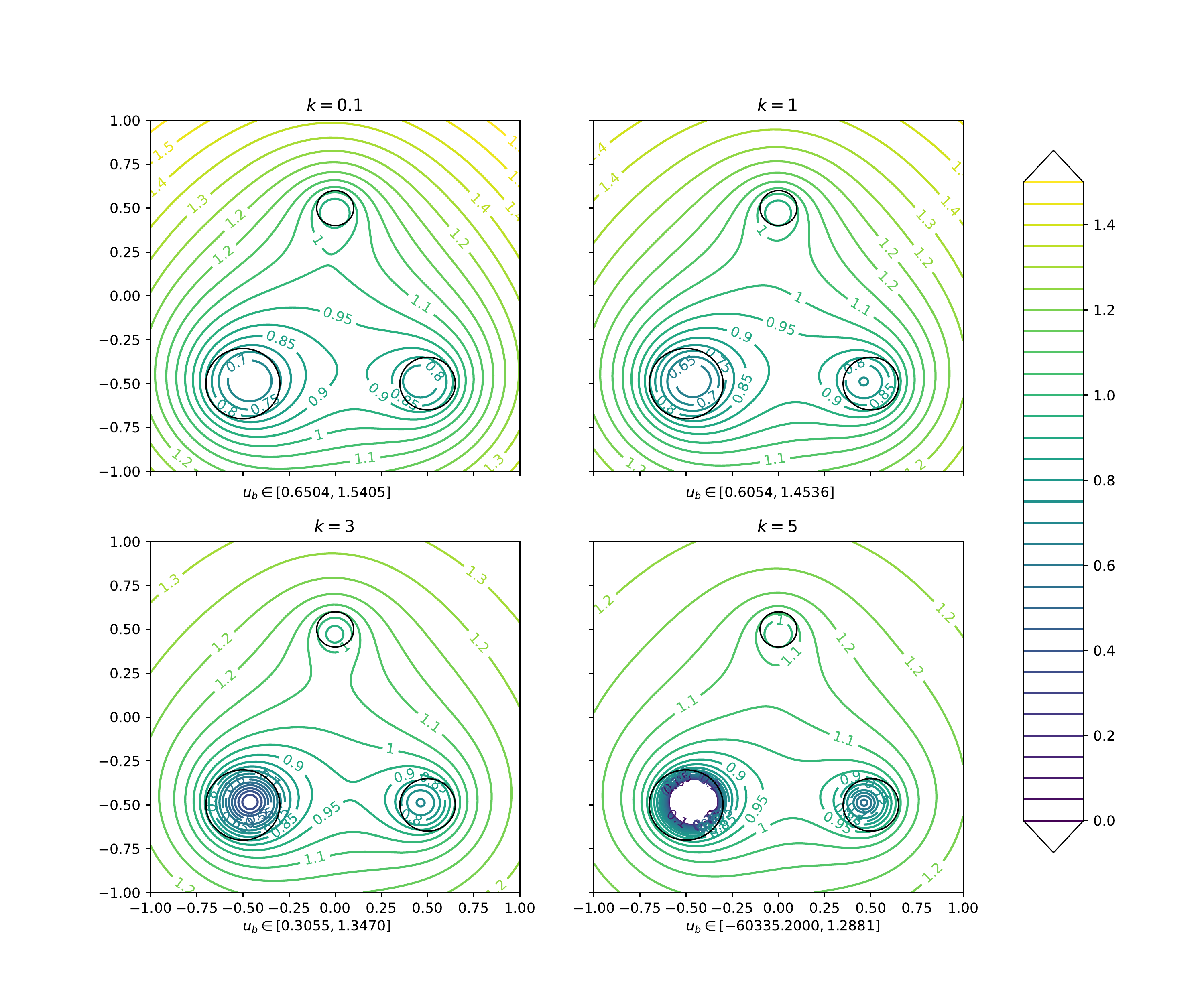}
	\caption{\textbf{Contour plot of the analytical solution $U_b$
	for three parallel tubes}. Cross-sectional perimeter shown by black circles.
	Plotted for tube radii $R = (0.2, 0.15, 0.1)$, $\varrho_i = R_i$,
	and various exponential diffusion coefficient functions (\cref{fig:Dbr}, $D_0 = 0.5$).
	The higher $k$ the larger the gradients at the tube-bulk interface (denser contour lines).
	The scale is cut off at $u_b = 0$ (hence the white hole in the biggest tube for $k=5$). The minimum and maximum values of $u_b$ are given under each plot.
	Note that due to the steep gradients in the case $k=5$ the minimum $u_b$ is found to be \num{-6e4}.
	We therefore consider this case as a hard-to-solve test case. We note that by increasing
	the kernel radius $\varrho$, $u_b$ is much smoother, cf. \cref{fig:roots_2R}. The interface reconstruction algorithm
	needs to reconstruct the average $u_b$ on the black line (tube-bulk interface) shown in this plot, regardless of $\varrho$.
	}
	\label{fig:roots3}
\end{figure}
In this section, we derive for the verification of the introduced method, two types of analytical solutions based on the
superposition of point source solutions.
We consider an infinite two-dimensional domain $\Omega \subset \mathbb{R}^2$ that cuts through $N$ parallel non-overlapping circular tubes
of different radii $R_i$. The flow resistance that the tubes pose to flow through $\Omega$ and its volume are neglected
so that $\Omega$ can be extended inside the tube radius and comprises the entire plane without circular cut-outs.

The error involved with this assumption has been numerically analyzed in \cite{Koch2021a} and found to be small.
The assumption is also commonly used in 1d-3d models
\cite{Peaceman1978,Hsu1989green,d2007multiscale,DAngelo2012,cattaneo2014computational,koeppl2018,Gjerde2018,Koch2019a}
based on the underlying assumption that tube radii are small.
As noted in the beginning of this section, an analytical solution for $\psi$ can
be obtained by superposition, see \cref{eq:psisuperpos}. The analytical solution
for $u_b$ is then found by numerical or exact inversion of the Kirchhoff transformation.

In the following, we compute the coefficients of such solutions numerically for given tube center positions $\vec{x}_i$,
tube radii $R_i$, and $u_{e,i}$, fixed $\gamma_i$, given $D_b(u_b)$, and $\varrho_i \geq R_i$. For a fully determined analytical
solution we require $N$ average interface unknowns $\hat{u}_{b,i}^\bigcirc$,
and some constant $C_\psi$ (corresponds to the choice $H\equiv C_\psi$ in \cref{eq:psisuperpos}). Furthermore, we have $N$ equations
\begin{equation}
	\label{eq:num_analytic_res}
	\hat{u}_{b,i}^\bigcirc = \frac{1}{|\per_i|}\int_{\per_i} \! T^{-1}(\psi(\vec{x})) \,\text{d}\vec{x} \approx \sum\limits_{k=1}^{K_\text{ip}} T^{-1}(\psi(\vec{x}_{i,k})) w_{i,k},
\end{equation}
where $\vec{x}_{i,k} \in \mathbb{R}^2$, $w_{i,k} \in \mathbb{R}^+$ are $K_\text{ip}$ integration points and weights.
We choose $\vec{x}_{i,k}$ to be uniformly distributed on $\per_i$.
To compute the solution numerically, we choose $\hat{u}_{b,1}^\bigcirc$ and then solve \cref{eq:num_analytic_res}
with a Newton method, where in every step the dense linear system
\begin{equation}
	\begin{bmatrix}
		\frac{\partial r_1}{\partial C_\psi} & \frac{\partial r_1}{\partial {u}_1} & \cdots & \frac{\partial r_1}{\partial {u}_N} \\
		\frac{\partial r_2}{\partial C_\psi} & \frac{\partial r_2}{\partial {u}_1} \\
		\vdots & & \ddots\\
		\frac{\partial r_N}{\partial C_\psi}  & & &\frac{\partial r_N}{\partial {u}_N} \\
	\end{bmatrix}
	\begin{bmatrix}
		\Delta C_\psi \\
		\Delta {u}_2 \\
		\vdots \\
		\Delta {u}_N \\
	\end{bmatrix}
	=
	\begin{bmatrix}
		r_1 \\
		r_2 \\
		\vdots \\
		r_N \\
	\end{bmatrix},
\end{equation}
is solved, where $u_i$ is short notation for $\hat{u}_{b,i}^\bigcirc$ and
\begin{equation}
	r_i = \hat{u}_{b,i}^\bigcirc - \sum\limits_{k=1}^{K_\text{ip}} T^{-1}(\psi(\vec{x}_{i,k})) w_{i,k} = 0,
\end{equation}
are the nonlinear residuals. The partial derivatives are approximated by numerical differentiation.
The resulting solution is denoted as $U_b$.

In a variation of the above algorithm we use
\begin{equation}
	\tilde{r}_i = \tilde{u}_{b,i}^\bigcirc - T^{-1}\left( \sum\limits_{k=1}^{K_\text{ip}} \psi(\vec{x}_{i,k}) w_{i,k} \right) = 0,
\end{equation}
and the resulting solution is denoted as $\tilde{U}_b$.
This variation exactly corresponds to the approximation \labelcref{eq:main_assumption_mvt}.
We will show in the subsequent numerical experiments that a distributed source scheme with an interface reconstruction
based on \cref{eq:nleqifpressure-mult} in a setup corresponding
to the two-dimensional parallel tube setup converges to the modified solution $\tilde{U}_b$.
However, $U_b$ and $\tilde{U}_b$ are very similar so that the error $\vert\vert U_b - \tilde{U}_{b,h} \vert\vert$
for a numerical approximation $\tilde{U}_{b,h}$ is small in practice.
Examples of $U_b$ for three parallel tubes are shown \cref{fig:roots3}.

\subsection{Discrete error measures}
To quantify the discretization errors, we define the following relative discrete $L^2$-errors for the unknown $u_b$,
its transformed variable $\psi = T(u_b)$, and the source term $q$ as
\begin{equation}
	E_{u_b} = \frac{1}{u_{b, \text{ref}}}\left[\frac{1}{\lvert \Omega_h \rvert}\sum_{K_\Omega \in \Omega_h} \lvert K_\Omega \rvert \left(u_{b, K_\Omega} - U_{b, K_\Omega} \right)^2\right]^{\frac{1}{2}},
\end{equation}
where $u_{b,\text{ref}}$ is a constant reference value chosen as $1$ (unless otherwise indicated) and $u_{b, K_\Omega}$ and $U_{b, K_\Omega}$ are the numerical and analytical solutions evaluated at the center of $K_\Omega$;
\begin{equation}
	E_{\psi} = \frac{1}{\psi_\text{ref}}\left[\frac{1}{\lvert \Omega_h \rvert}\sum_{K_\Omega \in \Omega_h} \lvert K_\Omega \rvert \left(\psi_{K_\Omega}-\Psi_{K_\Omega} \right)^2\right]^{\frac{1}{2}},
\end{equation}
where $\psi_\text{ref}$ is chosen as $0.1$ (unless otherwise indicated), and
\begin{equation}
	E_q = \frac{1} {q_\text{ref}} \left[ \frac{1}{ \lvert \Lambda_h\rvert} \sum_{K_\Lambda \in \Lambda_h} \left( q_{K_\Lambda}- Q_{K_\Lambda} \right) ^2\right]^{\frac{1}{2}},
\end{equation}
where $q_{K_\Lambda}$ and $Q_{K_\Lambda}$ are the numerical and the exact source
for the tube segment $K_\Lambda$, defined as the integral of $q$ in \cref{eq:nonlineardiffusion}
over $K_\Lambda$ and
$q_\text{ref} = \max\limits_{K_\Lambda \in \Lambda_h}\vert Q_{K_\Lambda}\vert$.
Finally, we analogously define relative discrete $L^2$-errors with respect to the modified analytical
solution $\tilde{U}$ (see \cref{sec:analytical_multiple}), and accordingly denote them as $\tilde{E}_{u_b}$, $\tilde{E}_{\psi}$, $\tilde{E}_q$.
We note that all errors are reported against the regularized solution $u_b$ rather than
the line source solution $\hat{u}_b$. The exact solutions for $u_e$ and $q$ are equivalent in both settings.

\subsection{Grid convergence tests}
In the following, we present grid convergence tests against the analytical solution
for a single tube and against the analytical solutions
of \cref{sec:analytical_multiple} for multiple parallel tubes.
The diffusion coefficient $D_b(u_b)$ is chosen as exponential function shown in \cref{fig:Dbr} with $D_0 = 0.5$.
Analytical expressions for the Kirchhoff transformation and its inverse are given in \cref{sec:app:exp}.
The diffusive wall permeability $\gamma$ is chosen as $1$.

As motivated in \cref{sec:analytical_multiple}, we expect that the numerical solution converges to
the modified solution $\tilde{U}_b$. However, $U_b$ and $\tilde{U}_b$ are expected to be very similar.
They are identical for a single infinite tube, where the approximation \labelcref{eq:main_assumption_mvt}
is exact.

\subsubsection{Single tube convergence rates}
\label{sec:convergence}
\begin{figure}[htb!]
	\centering
	\includegraphics[width=\textwidth]{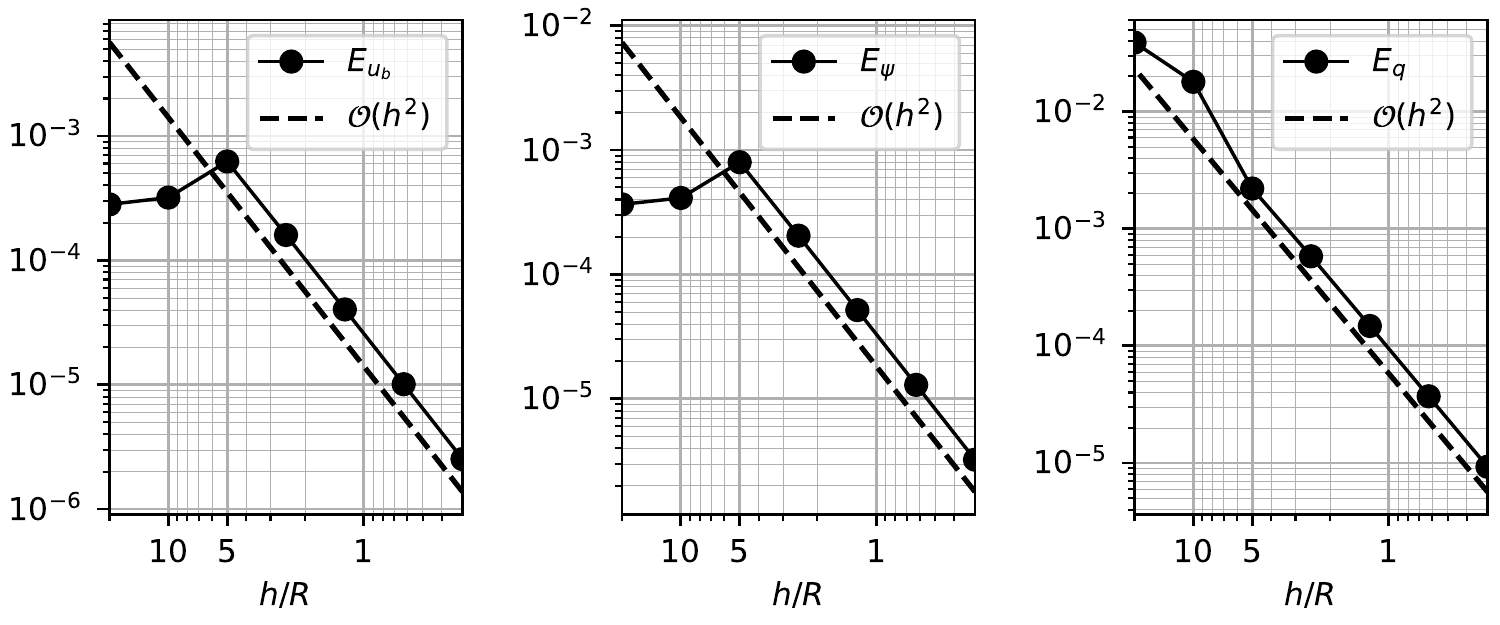}
	\caption{\textbf{Grid convergence of relative discrete errors for single infinite tube.}
	The kernel radius is $\varrho=5R$ and the tube radius $R=0.01$. $h$ is the discretization length.
	Following a pre-asymptotic range, second order convergence in all variables is observed for $h \leq \varrho$.}
	\label{fig:steady1conv}
\end{figure}
Let us consider \cref{eq:nonlineardiffusion} for an infinite straight tube
with constant $u_e$ embedded in an infinite bulk domain. The problem reduces to solving \cref{eq:polarcoodinatenonlineardiffusion}.
The analytical solution is given by \cref{eq:analytical_radsym}.
Note that for this radially symmetric case \cref{eq:main_assumption_mvt} is exact.
We therefore only report errors with respect to $U_b$ since $U_b$ and $\tilde{U}_b$ (as described in \cref{sec:analytical_multiple})
are identical for the single tube case.
We choose $R=0.01$, $\varrho=0.05$, $\hat{u}_b^\bigcirc = 0.5$, $u_e = 0.1$, and
numerically solve \cref{eq:polarcoodinatenonlineardiffusion} on the unit interval
with the analytical solution prescribed as Dirichlet boundary conditions at $r=1$.
The source term is computed based on $u_e$ using the proposed interface reconstruction, \cref{eq:nleqifpressure}.

Grid convergence results are shown in \cref{fig:steady1conv}.
The grid is uniformly refined, starting from $h=20R$. Initially, we observe a
pre-asymptotic range since the kernel support is not resolved by the grid yet.
After the 3rd refinement, where $h = \varrho$, all
errors decay quadratically with uniform grid refinement for all quantities,
the primary variable $u_b$, the transformed variable $\psi$, and the numerical source term $q$.
As observed by \citep{Koch2019a}, the onset of second-order convergence is
determined by the kernel radius rather than the tube radius. This allows for good
control of the error even for simulations where fine grids are not feasible.

\subsubsection{Multiple parallel tubes}
\label{sec:convergence_parallel}
Next, we consider a two-dimensional domain
$\Omega =[-1, 1] \times [-1, 1]$ that perpendicularly cuts three tubes
with radii $(R_\text{max}, \frac{3}{4}R_\text{max}, \frac{1}{2}R_\text{max})$
centered at $\vec{x}_1 = (-0.5, -0.5)$, $\vec{x_2} = (0.5, -0.5)$, $\vec{x_3} = (0, 0.5)$, respectively.
The tube unknowns $u_{e,i}$ are given by $(0.3, 0.2, 0.1)$ and the kernel radii are $(0.4, 0.3, 0.2)$, respectively.
The average interface unknown of the largest tube $\hat{u}_{b,1}^\bigcirc$ is fixed as $0.8$.
The analytical solutions $U_b$ and $\tilde{U}_b$ are computed as described in \cref{sec:analytical_multiple}.
The solution $U_b$ is shown for different $k$ in \cref{fig:roots3} for $\varrho_i = R_i$. The same
solution for $\varrho_i = 2R_i$ is shown in \cref{sec:app:tworho}.

In the following, we investigate the influence of the exponential rate parameter $k$ of $D_{b}(u_b)$
and the influences of the tube radii $R_\text{max}$
on the discretization and the error involved in approximation \cref{eq:main_assumption_mvt}.
We compute grid convergence both against $U_b$ and $\tilde{U}_b$.
The mesh $\Omega_h$ is uniformly refined starting with $4\times 4$ cells.
On boundaries, we enforce the respective analytical solution
as Dirichlet boundary condition for $u_b$.

\begin{figure}[htb!]
	\centering
	 \includegraphics[width=\textwidth]{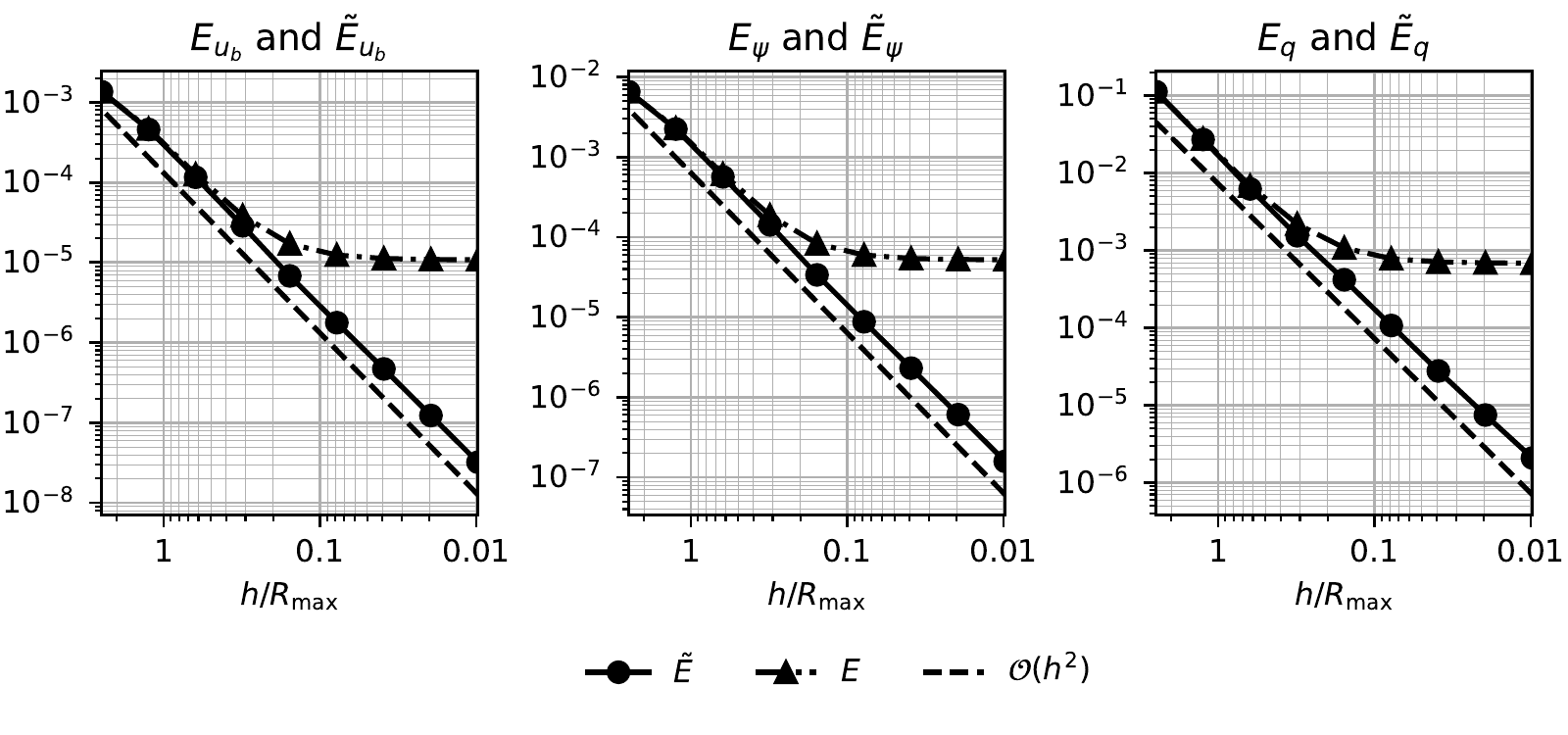}
	\caption{\textbf{Grid convergence for relative discrete errors $E$ and $\tilde{E}$}.
	The maximum tube radius is $R_\text{max}=0.2$ (the three tube radii are $0.2, 0.15, 0.1$),
	kernel radii are $\varrho_i = 2R_i$ and the exponential rate parameter is $k = 1$.}
	\label{fig:epepsieqformutipletubes}
\end{figure}

In the first case, we set $k=1$ and $R_\text{max}=0.2$.
The errors $E_{u_b}$, $E_{\psi}$ and $E_q$ are shown in \cref{fig:epepsieqformutipletubes}
with uniform grid refinement.
As motivated in \cref{sec:analytical_multiple}, we see convergence to the modified analytical solution $\tilde{U}_b$.
Second-order convergence is observed for all relevant quantities.
For the convergence test against $U_b$, we observe a non-reducible error.
Due to the way the analytical solution is constructed,
we can identify this error as the model error caused by approximation \labelcref{eq:main_assumption_mvt}.
However, most interestingly, this error is very small (less than \SI{0.1}{\percent} for $\tilde{E}_q$) in this case and shows only an influence
in the convergence plot for grid discretization length below the tube radius, i.e. $h < R_\text{max}$.

\begin{figure}[htb!]
	\centering
	\includegraphics[width=\textwidth]{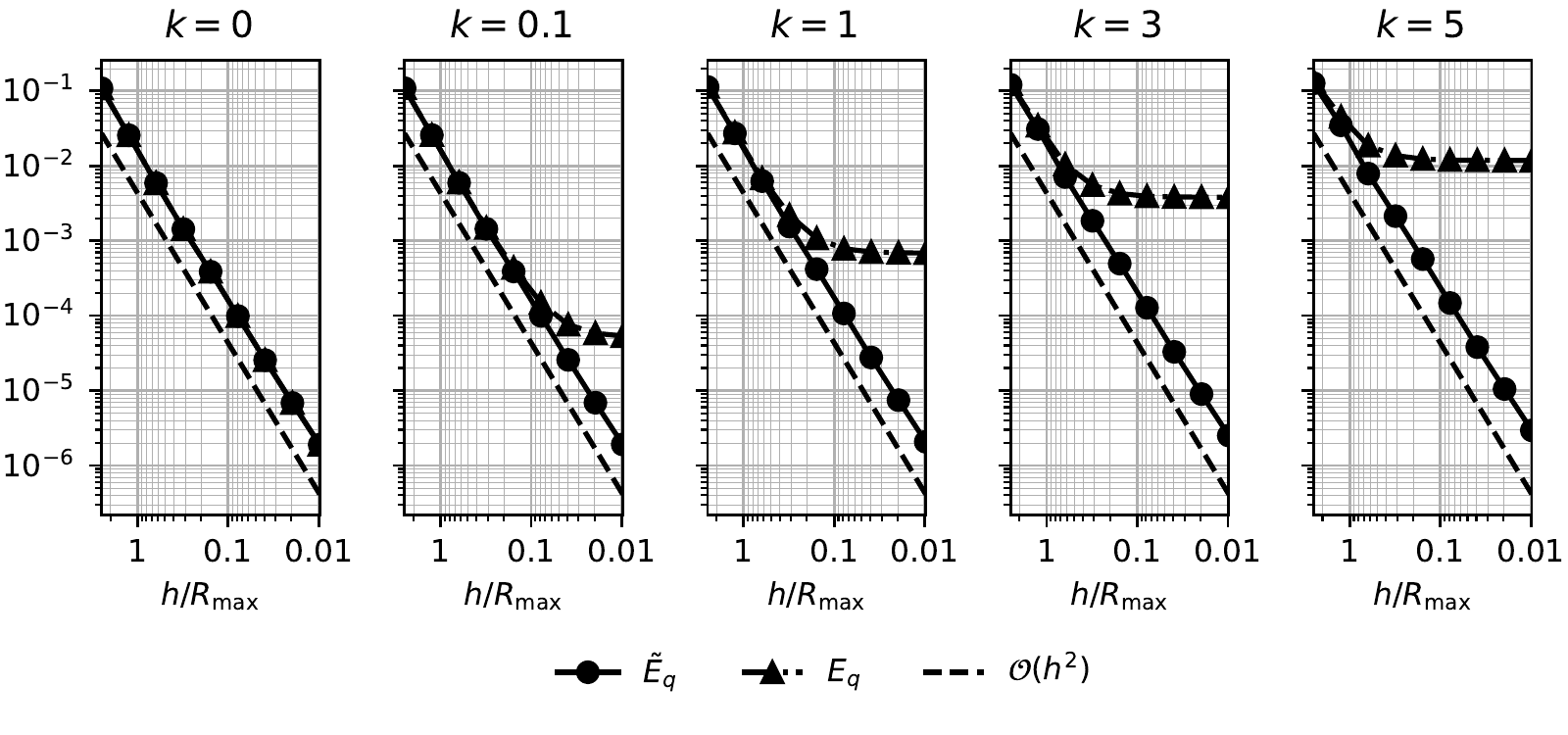}
	\caption{\textbf{Grid convergence of source errors for different exponential coefficients $k$.}
	Source errors $E_q$ and $\tilde{E}_q$ for $k = 0.1, 1, 3$ and $5$ with $R = (0.2, 0.15, 0.1)$.}
	\label{fig:kformultipletubes}
\end{figure}

In the second case,
we investigate more challenging cases by varying $k$, cf.~\cref{fig:Dbr}.
The source error for different $k$ is shown in \cref{fig:kformultipletubes}.
As convergence of all three fields
$u_b$, $\psi$, $q$ are confirmed in \cref{fig:steady1conv,fig:epepsieqformutipletubes},
we only report errors for $q$ in the following, allowing for a more concise presentation.

For $k=0$ the exponential function reduces to a constant. Therefore,
this case corresponds to the linear stationary diffusion equation analyzed in~\citep{Koch2019a}.
Due to the mean value theorem for harmonic functions, approximation \labelcref{eq:main_assumption_mvt}
is exact and $U_b = \tilde{U}_b$. The larger $k$ the stronger the changes in $D_b$.
In particular, as shown in \cref{fig:roots3}, the gradient of $u_b$ at the tube interface
is large for larger $k$ and the value of $u_b$ varies considerably along the tube perimeter.
Nevertheless, the largest observed error for $k=5$ is ${E}_q \approx \SI{1}{\percent}$,
and only dominates the discretization error for $h < R_\text{max}$.

\begin{figure}[htb!]
	\centering
	\includegraphics[width=\textwidth]{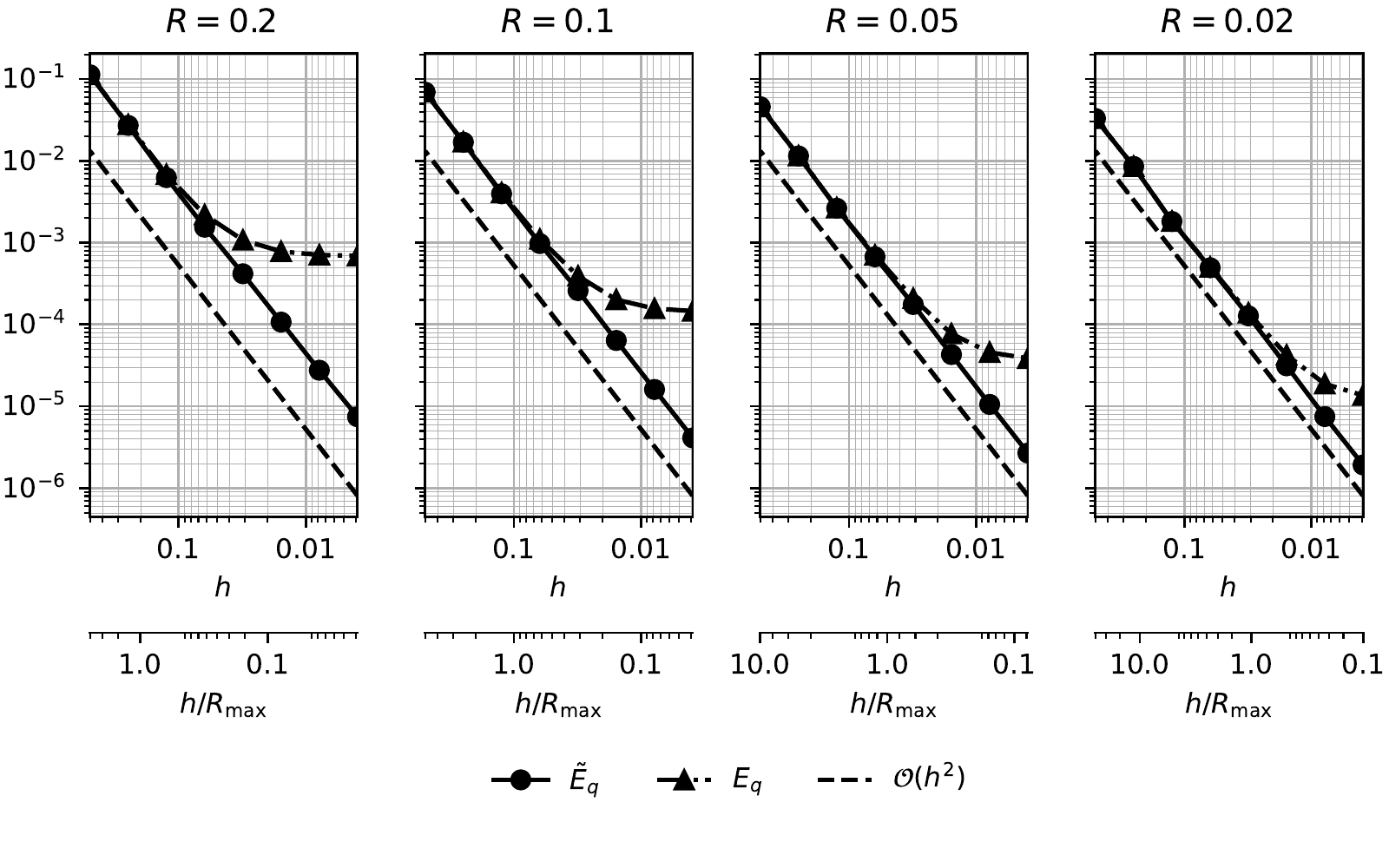}
	\caption{\textbf{Grid convergence of source errors for different tube radii $R$.}
	Source errors $E_q$ and $\tilde{E}_q$ with $k = 1$ and $R_\mathrm{max} = 0.2, 0.1, 0.05, 0.02$.}
	\label{fig:rformultipletubes}
\end{figure}
In a third case, we fix $k=1$ and vary the tube radii $R_\mathrm{max} =0.2, 0.1, 0.05$ and $0.02$.
The analytical solution is computed such that the source term of the largest tube is equivalent for all cases.
The source errors are presented in \cref{fig:rformultipletubes}.
It is evident that the error due to approximation \labelcref{eq:main_assumption_mvt}
decreases with smaller tube radius. This is in good agreement with
the error estimate in \cref{sec:errorestimate}. In particular, we can see again
that the error is only relevant in comparison to the discretization error for $h < R_\mathrm{max}$.

\subsubsection{Influence of the kernel radius $\varrho$}
\begin{figure}[htb!]
	\centering
	\includegraphics[width=\textwidth]{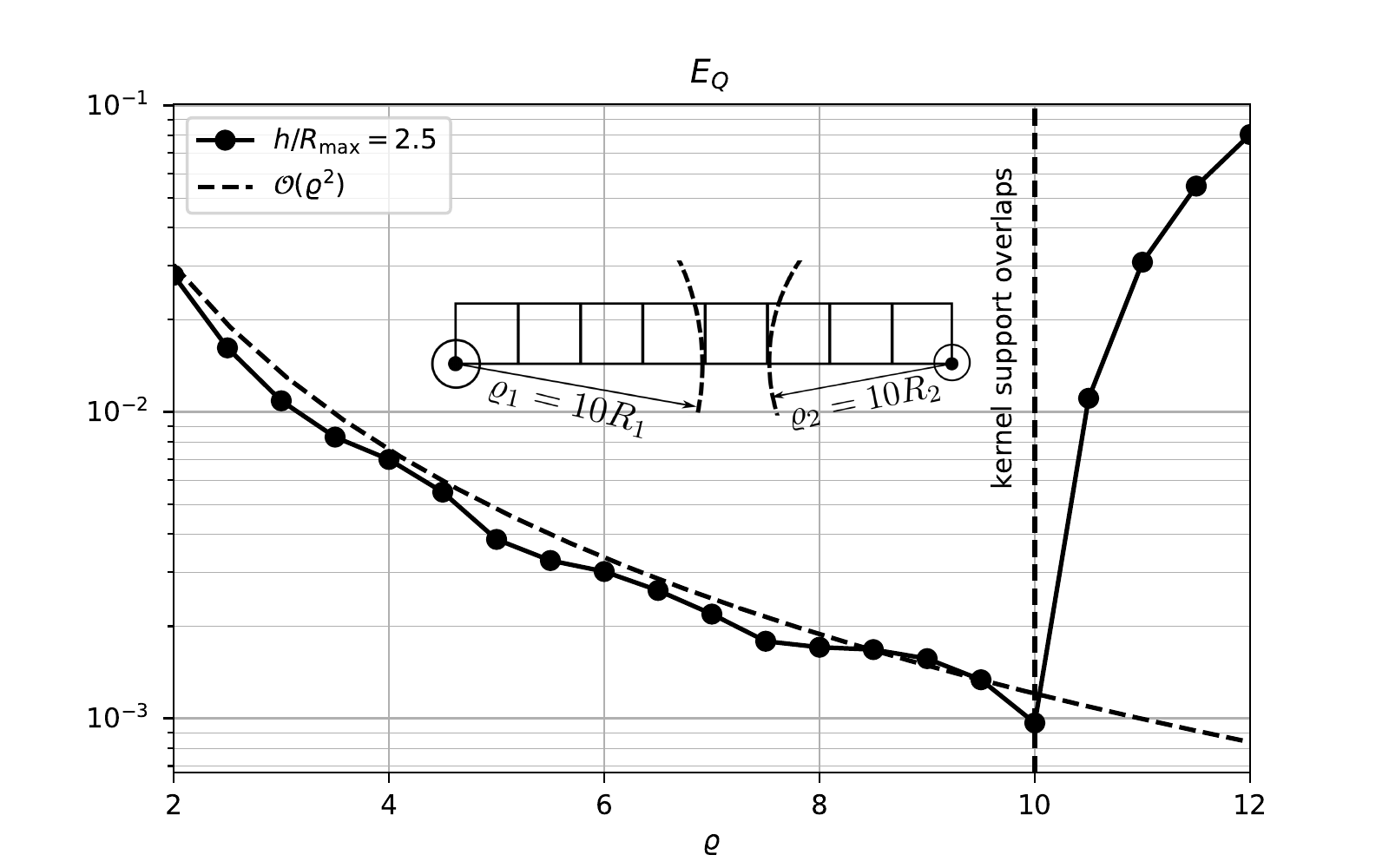}
	\caption{\textbf{Source errors $E_q$ for different kernel radii $\varrho$.}
	The maximum tube radius is $R_\mathrm{max} =0.2$ and $k=1$. Semi-logarithmic plot.
	Exact source terms $Q_i$ are the same for all $\varrho$.}
	\label{fig:qoverrho}
\end{figure}
In the fourth test case, we investigate the influence of the kernel
radius $\varrho$ on the discretization error. Therefore, we choose
a case where the approximation error due to \cref{eq:main_assumption_mvt}
does not dominate the total model error.
We choose $R_\mathrm{max} = 0.05$ and $k = 1$. The discretization length $h$
is fixed at $0.125$. All other parameters are the same as in \cref{sec:convergence_parallel}.
The kernel radius is increased from $2R_i$ to $12R_i$.
The resulting source errors $E_q$ with respect to the analytical solution $U_b$
are shown in \cref{fig:qoverrho}. The error decays with increasing
kernel radius. Interestingly, the speed of this decay is comparable
with the error decay by grid refinement and matches the observations in \citep{Koch2019a}
for the linear diffusion equation. It can also be seen that as soon as the
kernel regions of the largest two tubes start to overlap the source error increases again.
The error decay is best explained by the better approximation of $u_b(\vec{x}_i)$ used
in the reconstruction algorithm, cf.~\cref{eq:nleqifpressure} and \cref{eq:nleqifpressure-mult}. Since $u_b$ is increasingly regularized with
increasing kernel support, it becomes easier to approximate the function numerically.

\subsubsection{Discrete evaluations of $u_{b,\delta}$ or $u_{b,0}$}
\label{sec:delta}
As noted in \cref{sec:reconstruction}, the reconstruction allows to consider
that the numerically measured quantity $u_{b,K_\Omega}$ in cell $K_\Omega$ does not represent the
value on the centerline but rather a value in some distance $\delta$. This
is particularly relevant for coarse discretizations where it may make a difference whether the root segment is located in the middle or the corner of a 3D cell.

As a fifth case, we present the results of the second case (where $u_{b,K_\Omega}$ is interpreted as $u_0$)
in comparison with results for which we assumed in the reconstruction that $u_{b,K_\Omega}$
represents the $u_{b,\delta}$ and $\delta$ is chosen as the mean minimum distance
between tube segment and bulk cell~\citep{Koch2018a},
\begin{equation}
\delta = \frac{1}{\vert K_\Omega \vert} \int_{K_\Omega} \min_{\boldsymbol{x'} \in K_\Lambda} \vert\vert \boldsymbol{x} - \boldsymbol{x'} \vert\vert_2 \,\text{d}x.
\end{equation}
\begin{figure}[htb!]
	\centering
	\includegraphics[width=\textwidth]{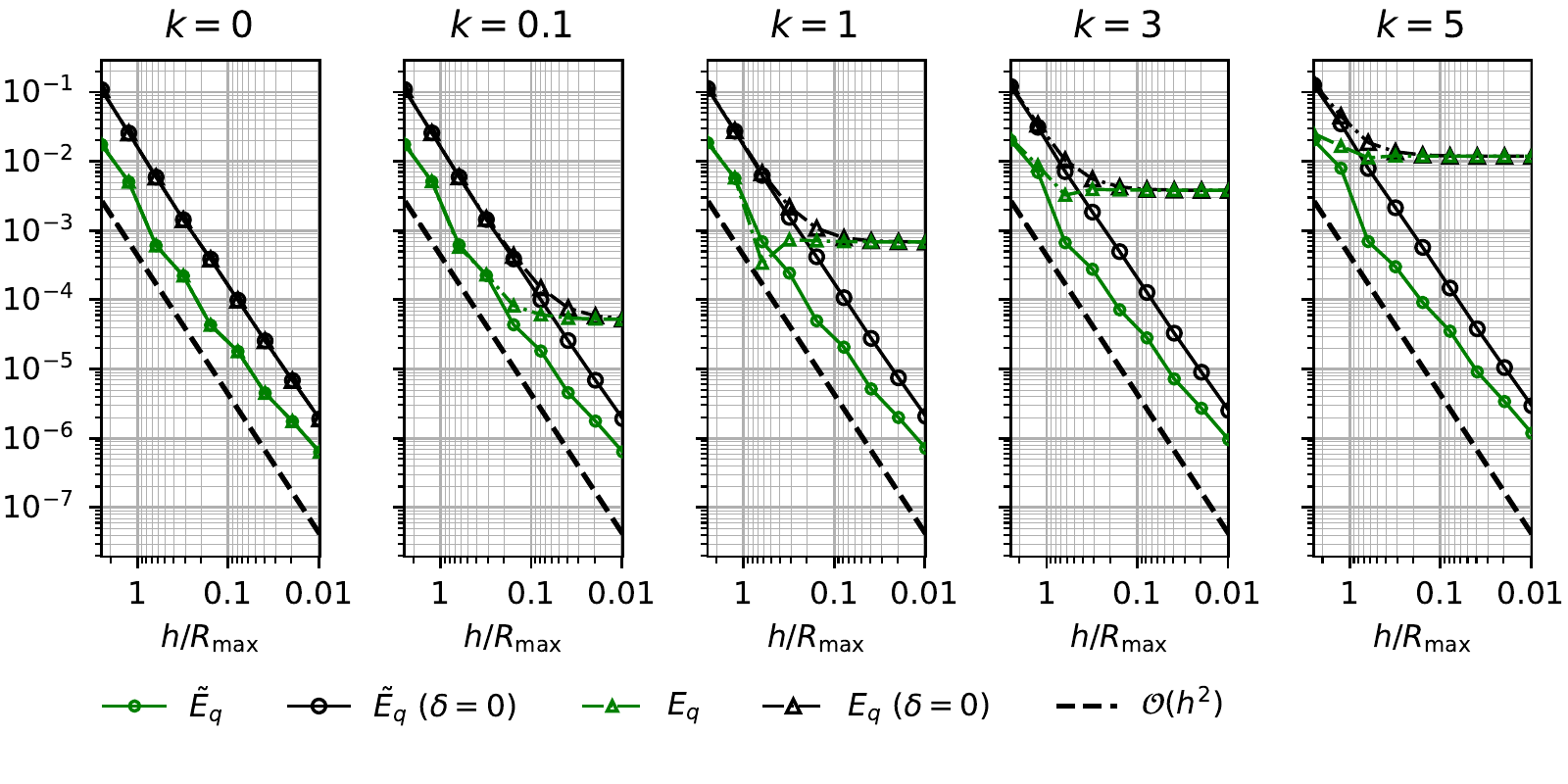}
	\caption{\textbf{Grid convergence of source errors with non-zero $\delta$.}
	Source errors $E_q$ and $\tilde{E}_q$ for different exponential coefficients $k=0.1, 1, 3$ and $5$ with $R = (0.2, 0.15, 0.1)$,
	where $\delta$ is defined as the mean distance between the tube and bulk cell containing the tube (in green).
	$\delta = 0$ (in black) is the situation where the tube is located at the center of the bulk cell.}
	\label{fig:kformultipletubesexpdelta}
\end{figure}
The resulting source errors are shown in \cref{fig:kformultipletubesexpdelta}.
We observe that the error is significantly reduced. As for the previous cases,
the error curve flattens as soon as the error is dominated
by the mean value approximation error.

\subsection{Root--soil interaction scenario}
\label{sec:rootnetwork}
In the following application scenario, we compute root water uptake with small root system architecture
obtained from MRI measurements. The scenario is similar to benchmark scenario C1.2 presented in \citep{Schnepf2019benchmark}.
However, we solve a stationary problem for various root collar pressures enforced as Dirichlet boundary
conditions at the root collar.

\begin{figure}[!htb]
	\centering
	\includegraphics[width=0.79\textwidth]{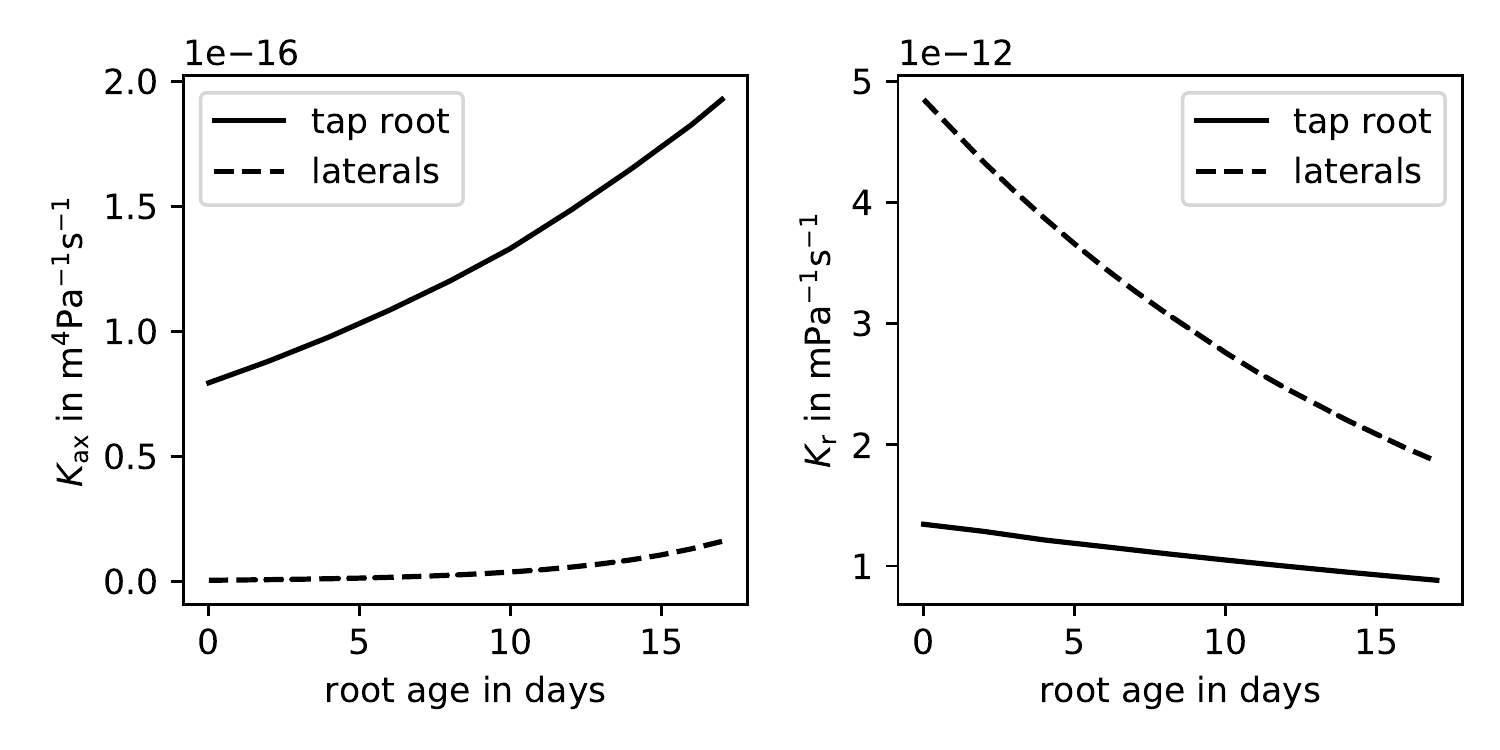}
	\includegraphics[width=0.20\textwidth]{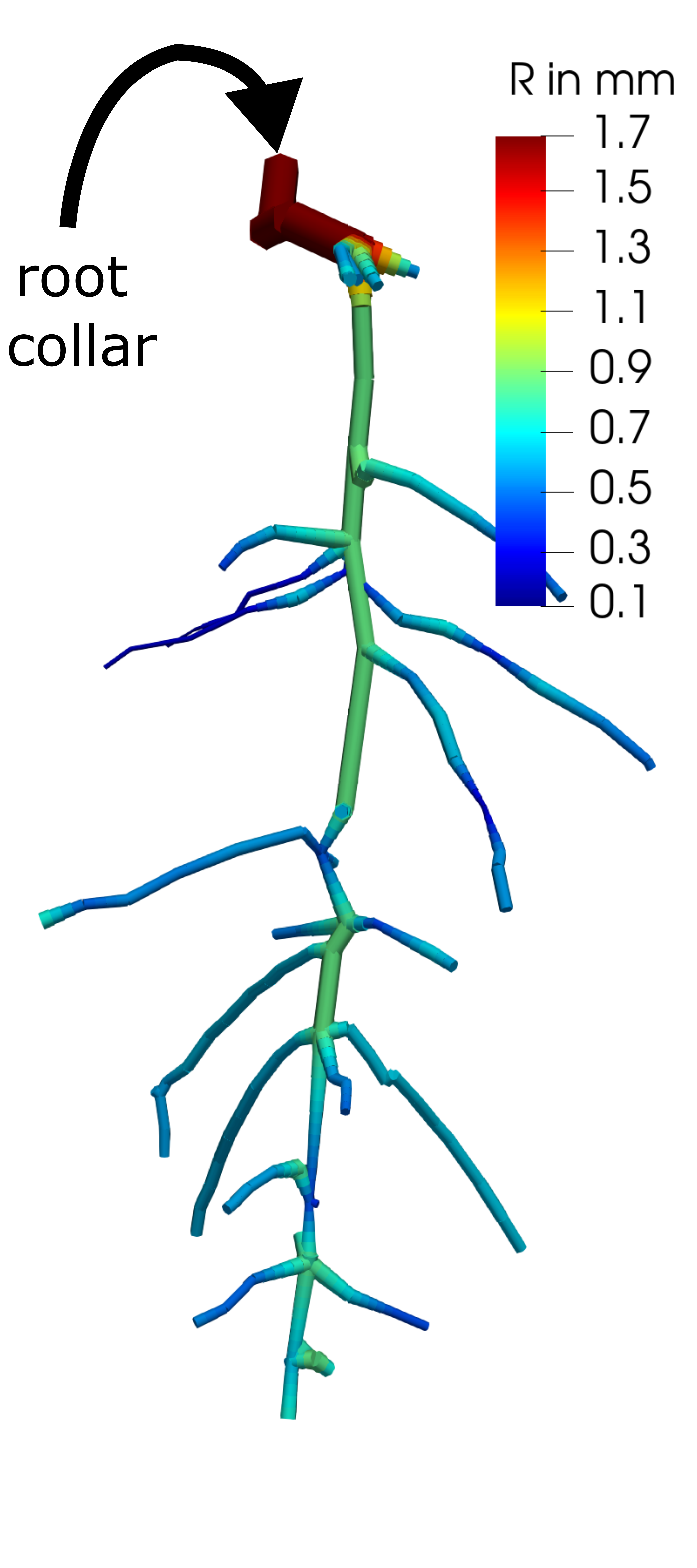}
	\caption[Root conductivity and 8-day-old lupin root system]{\textbf{Root conductivities and radius for a lupin root system.}
	Left and middle, age-dependent hydraulic root conductivities from~\citep{Schnepf2019benchmark}.
	The axial root conductivity $K_\text{ax}$ corresponds to $D_e$ and
	the radial root conductivity $K_\text{r}$ corresponds to $\gamma$ in the nonlinear diffusion equation.
	Right, 8-day-old lupin root system reconstructed from MRI data (courtesy of M. Landl, FZ Jülich).
	Grid data available from \citep{Koch2019BenchmarkC12Data}.
	The root segment radius is visualized to scale. The rooting depth is about \SI{10}{\centi\m}.
	Figure adapted from \citep{Koch2020PhDthesis}.}
	\label{fig:root-conductivities-geometry}
\end{figure}

To make it easier for readers familiar with root-soil interaction, we introduce several symbols and relate
them to the symbols in \cref{eq:nonlineardiffusion,eq:1d,eq:source}. The soil pressure $p_s$ and
root pressure $p_r$ (in \si{\pascal}) correspond to the unknowns $u_b$ and $u_e$.
The term $\mu^{-1}K k_r$ corresponds to $D_b$, where $\mu = \SI{1e-3}{\pascal\s}$ is the viscosity of water,
$K$ (in \si{\square\m}) is the intrinsic permeability of the solid matrix, and $k_r$ is the dimensionless relative permeability.
Relative permeability is commonly modeled as a nonlinear function of water content which in turn can be described by a nonlinear function of
soil water pressure. The axial root conductivity $K_\text{ax}$ (in \si{\m\tothe{4}\per\pascal\per\s}) corresponds to $D_e$,
and the radial root conductivity $K_\text{r}$ (in \si{\m\per\pascal\per\s}) corresponds to $\gamma$. We can then reformulate
\cref{eq:nonlineardiffusion,eq:1d,eq:source} to obtain a stationary root-water uptake model neglecting gravity,
\begin{align}
	\label{eq:nonlineardiffusion-root}
	-\div \left(\mu^{-1}k_r(p_s) K \grad p_s \right)  &= q\Phi_\Lambda & \text{in} \quad \Omega, \\
	\label{eq:1d-root}
	- {\partial_s}\left( K_\text{ax} {\partial_s} p_r \right) &= -q & \text{on} \quad \Lambda, \\
	\label{eq:source-root}
	q &= -2\pi R K_\text{r} (\hat{p}_s^\bigcirc - p_r ),
\end{align}
where $\hat{p}_s^\bigcirc$ denotes the average soil pressure at the root-soil interface.
\begin{figure}[htb!]
	\centering
	\includegraphics[width=\textwidth]{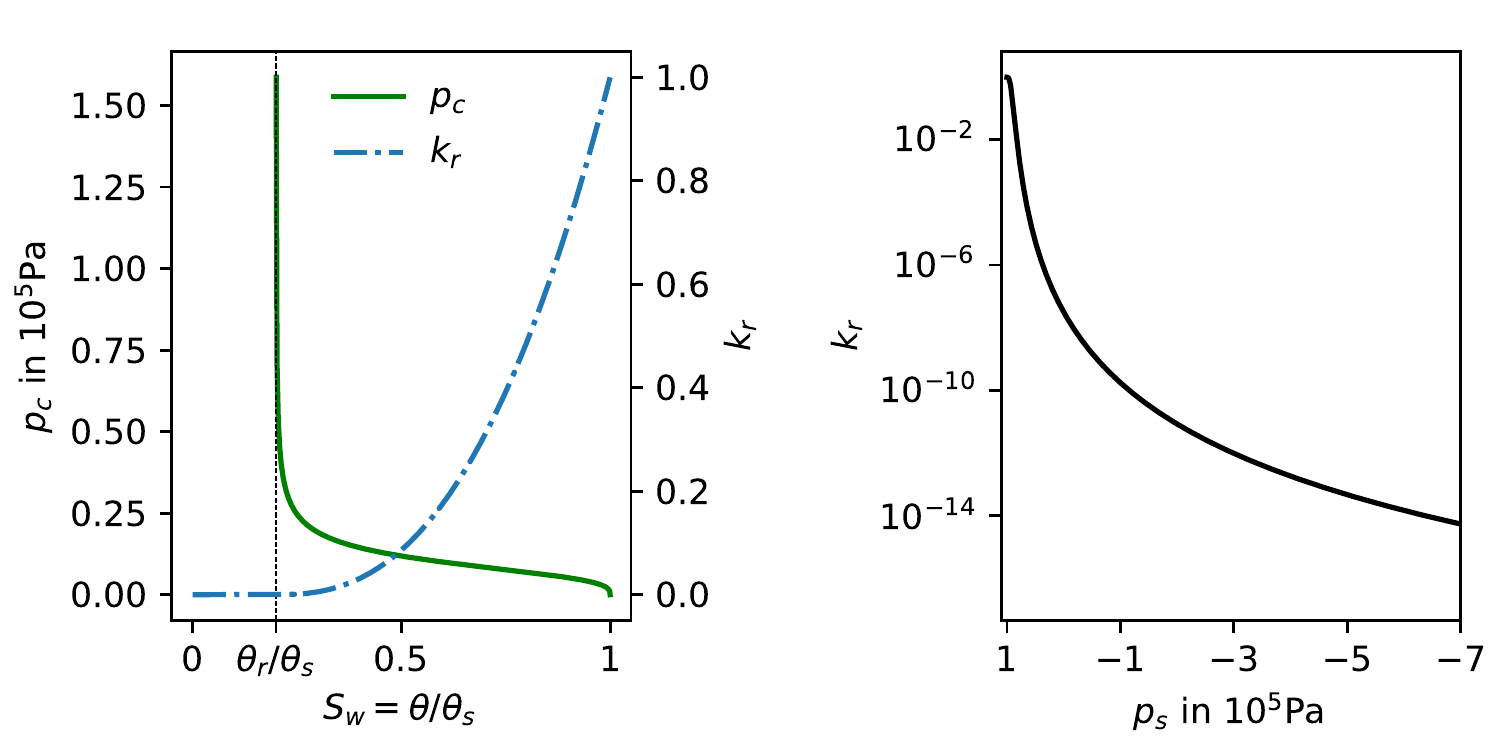}
	\caption{\textbf{Van Genuchten-Mualem model.} Left, relative permeability $k_r$ and capillary pressure $p_c$ over water content $\theta$
	using the Van Genuchten-Mualem model for a loamy soil, $K = \SI{5.89912e-13}{\square\m}$, $\theta_r = 0.08$, $\theta_s = 0.43$,
	$\alpha=\SI{4.077e-4}{\per\pascal}$, $n=1.6$. Right, the relative permeability $k_r$ as a function of soil water pressure $p_s$.
	The term $\mu^{-1} K k_r(p_s)$ in the root-soil scenario corresponds to $D_b(u_b)$ in the general nonlinear diffusion model.}
	\label{fig:krS_krPw}
\end{figure}

Let $\theta = S_w \theta_s$ denote the water content,
where $S_w$ is the water saturation and $\theta_s$ is the water content at saturation (equal to the porosity of the soil),
and let $\theta_r$ denote the residual water content.
The relative permeability is modeled by the Van Genuchten-Mualem model~\citep{mualem1976,van1980closed}
\begin{equation}
	\label{eq:vg}
	\begin{split}
	k_r(S_e) &= S_e ^{\lambda}[1 - (1 - S_e^{1/m})]^2, \\
	S_e(p_c) &= \left( (\alpha p_c)^n + 1 \right)^{-m}, \\
	p_c(p_s) &= - p_s,
	\end{split}
\end{equation}
where $p_c$ (in \si{\pascal}) is called capillary pressure, $S_e = \theta_r\theta_s^{-1}$ and $\theta_r$,
$\theta_s$, $\alpha$, $n$, $m = 1 - n^{-1}$ are material-dependent parameters.
In the following, we use a parametrization corresponding to loam
given in \citep{Schnepf2019benchmark}, see~\cref{fig:krS_krPw}.
The functions in \cref{eq:vg} are plotted in \cref{fig:krS_krPw}.
The axial and radial root conductivities vary along the roots dependent on the root age.
These root conductivity values are plotted in \cref{fig:root-conductivities-geometry}.
For tabularized values, we refer to \citep{Schnepf2019benchmark}.

The root system shown in \cref{fig:root-conductivities-geometry} is embedded in a box-shaped domain
with dimension $8\times8\times15$~\si{\centi\m}. The top of the box intersects with the root collar
at $x_3 = \SI{0}{\centi\m}$. The bottom of the domain is located at $x_3 = \SI{-15}{\centi\m}$. We
prescribe a water saturation of $S_w = 0.4$ (corresponding to $p_s = \SI{0.78e5}{\pascal}$) at all
sides except for the top boundary where we enforce a zero-flow Neumann boundary condition.
In the root domain, we prescribe no-flow boundary conditions at root tips and a fixed
pressure $p_{r,c}$ at the root collar. We solve the same scenario for
$p_{r,c} = \num{0.0}, \num{-0.5e5}, \num{-1.0e5}, \num{-2.5e5}$, and \SI{-5.0e5}{\pascal}. With decreasing root pressure,
the flow rate of water leaving the domain at the root collar (transpiration rate) increases
and the root-soil interface dries out. Dry soil (low water saturation) corresponds to
a strong decrease of the local hydraulic conductivity, cf.~\cref{fig:krS_krPw}.

We compare the results obtained with the presented distribution kernel-based method (DS)
for $\varrho_i = 3R_i$ and $\delta$ chosen as in \cref{sec:delta}
with two previously published methods. In \citep{Koch2021a} the root-soil
interface is fully resolved by a locally refined unstructured three-dimensional mesh. The roots are modeled with \cref{eq:1d-root} on a network of line segments. The solution $p_r$
(or $u_e$) is projected onto the closest surface on the three-dimensional mesh to
evaluate a source term similar to \cref{eq:source-root}. Since the root-soil interface
is resolved, this resolved-interface method (PS) does not suffer from the approximation errors
described in the previous sections, however, this comes with a higher computational cost as
we will demonstrate. Secondly, we compare with the method presented in \citep{koeppl2018}
and adapted for root-soil interaction as described in \citep{Koch2018a}. All methods are
implemented in \dumux~\citep{Dumux32019,Koch2018a} and are therefore easily comparable.
In fact they share most of the source code and mostly differ in the way the soil and root domain are coupled.

\begin{figure}[htb!]
	\centering
	\includegraphics[width=1.0\textwidth]{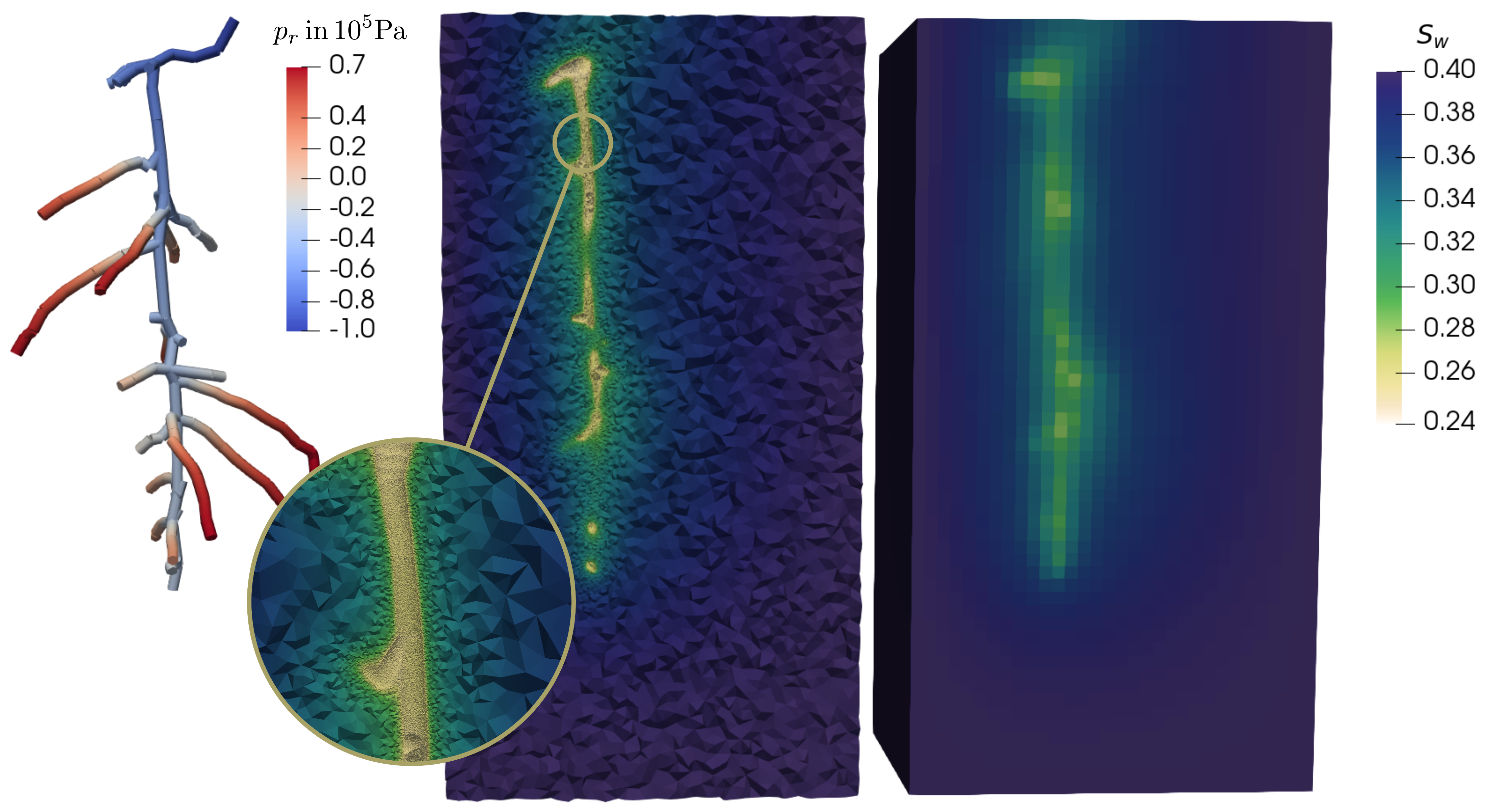}
	\caption{\textbf{Model comparison for root water uptake scenario.} Simulation result for $p_{r,c} = \SI{-1e5}{\pascal}$. Left, root pressure distribution
	in the root. Right, vertical cut through the soil domain (a box of $8\times8\times15$~\si{\centi\m}
	centered at $\vec{x} = (0, 0, -7.5)$~\si{\centi\m}) in the $x_2-x_3$-plane
	at $x_1 = \SI{0.7}{\centi\m}$. The left cut shows the soil saturation for a projection-based
	mixed-dimension method with fully resolved root-soil interface [PS]. The computational mesh has
	$12$M cells and the average cell diameter of the smallest \SI{10}{\percent} of cells
	is $\overline{h}_{10} = \SI{1.6}{\micro\m}$. The right cut
	shows the corresponding saturation distribution for the new kernel-based mixed-dimension scheme [DS] on
	a mesh with $60$k cells and $h_\text{min} = \SI{0.25}{\centi\m}$.}
	\label{fig:rootsoil1}
\end{figure}
The simulation result for $p_{r,c} = \SI{-1e5}{\pascal}$ and both methods PS and DS is shown in \cref{fig:rootsoil1}.
A close-up shows the locally refined grid necessary to accurately resolve the root-soil interface.
Due to the regularizing effect of both the distribution kernel and the coarse grid the DS solution $u_b$
does not contain the low saturation values found on the interface in the PS solution. Hence, the question is
if these values can be accurately reconstructed from $u_b$. As a global measure of how accurate
the source terms $q$ are approximated, we compute the transpiration rate at the root collar. Due
to mass conservation, the transpiration rate is given by
\begin{equation}
r_T = \int_\Lambda q(s) \,\text{d}s.
\end{equation}

\begin{figure}[htb!]
	\centering
	\includegraphics[width=1.0\textwidth]{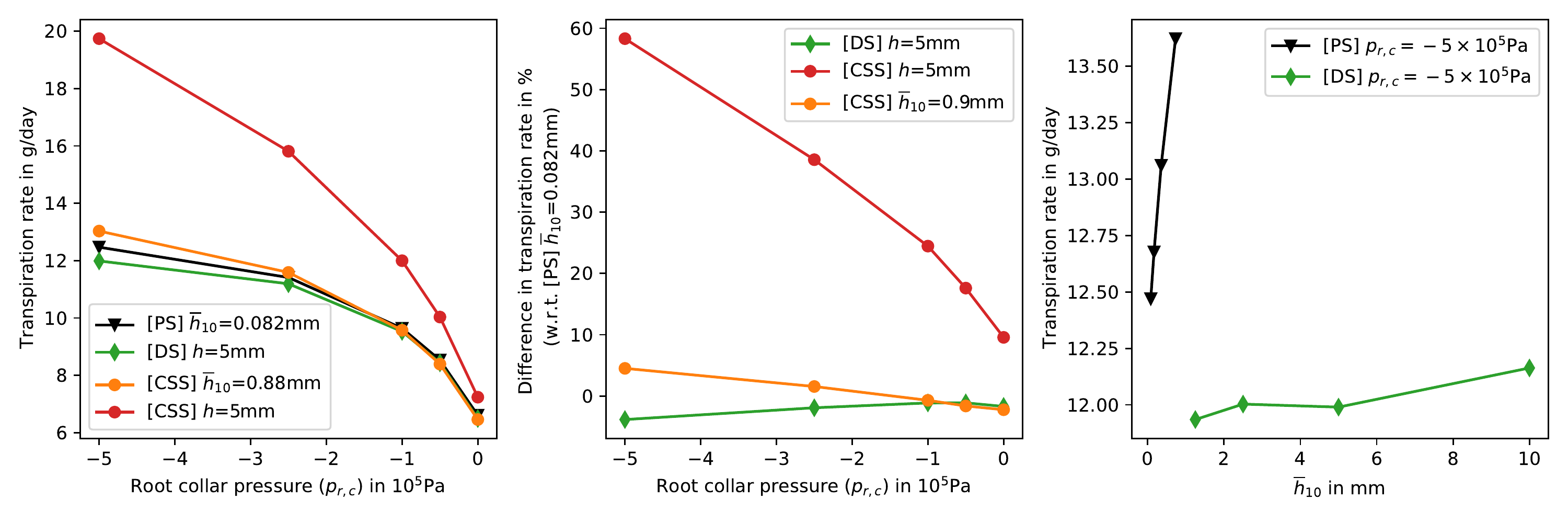}
	\caption{\textbf{Predicted transpiration rates for different grids and methods.} [DS] distributed source method (this paper),
	[CSS] cylinder surface method, cf.~\citep{koeppl2018,Koch2018a},
	[PS] projection method, resolved root-soil interface, cf.~\citep{Koch2021a}.
	The symbol $\overline{h}_{10}$ denotes the average diameter of the smallest \SI{10}{\percent} of the cells.
	The DS method uses a uniform mesh with cell diameter $h \equiv \overline{h}_{10}$. The coarse mesh for the CSS method is the
	same uniform mesh as for DS. The fine mesh for CSS is a locally refined mesh.}
	\label{fig:transpiration}
\end{figure}
Transpiration rates for all three methods and all root collar pressures are shown
in \cref{fig:transpiration}. Firstly, it can be observed that the DS method approximates
all transpiration rates with a maximum relative difference of \SI{3}{\percent}
to the high fidelity PS solution, notably using a quite coarse grid ($h = \SI{5}{\milli\m}$).
Using the same grid resolution the CSS method shows a difference of
\SI{10}{\percent} for $p_{r,c} = \SI{0}{\pascal}$ and even \SI{60}{\percent} for
$p_{r,c} = \SI{-5e5}{\pascal}$. Since CSS is consistent the results improve with grid refinement.
The difference to the reference solution is comparable to that of DS in terms of the
transpiration rate using strong local
grid refinement resulting in a mesh of $1.9$M cells.
Although the finest mesh used for the PS method in this work is locally refined with a total
of $12$M cells and well-resolved root-soil interface (cf. \cref{fig:rootsoil1}),
the grid convergence results shown in the right-most plot in \cref{fig:transpiration}
suggest that the transpiration rate is not fully converged yet.
Interestingly, following the trend, the transpiration rate is expected to get even closer
to the solution of the DS method which is already stable with grid refinement for rather coarse grids.

\begin{figure}[htb!]
	\centering
	\includegraphics[width=1.0\textwidth]{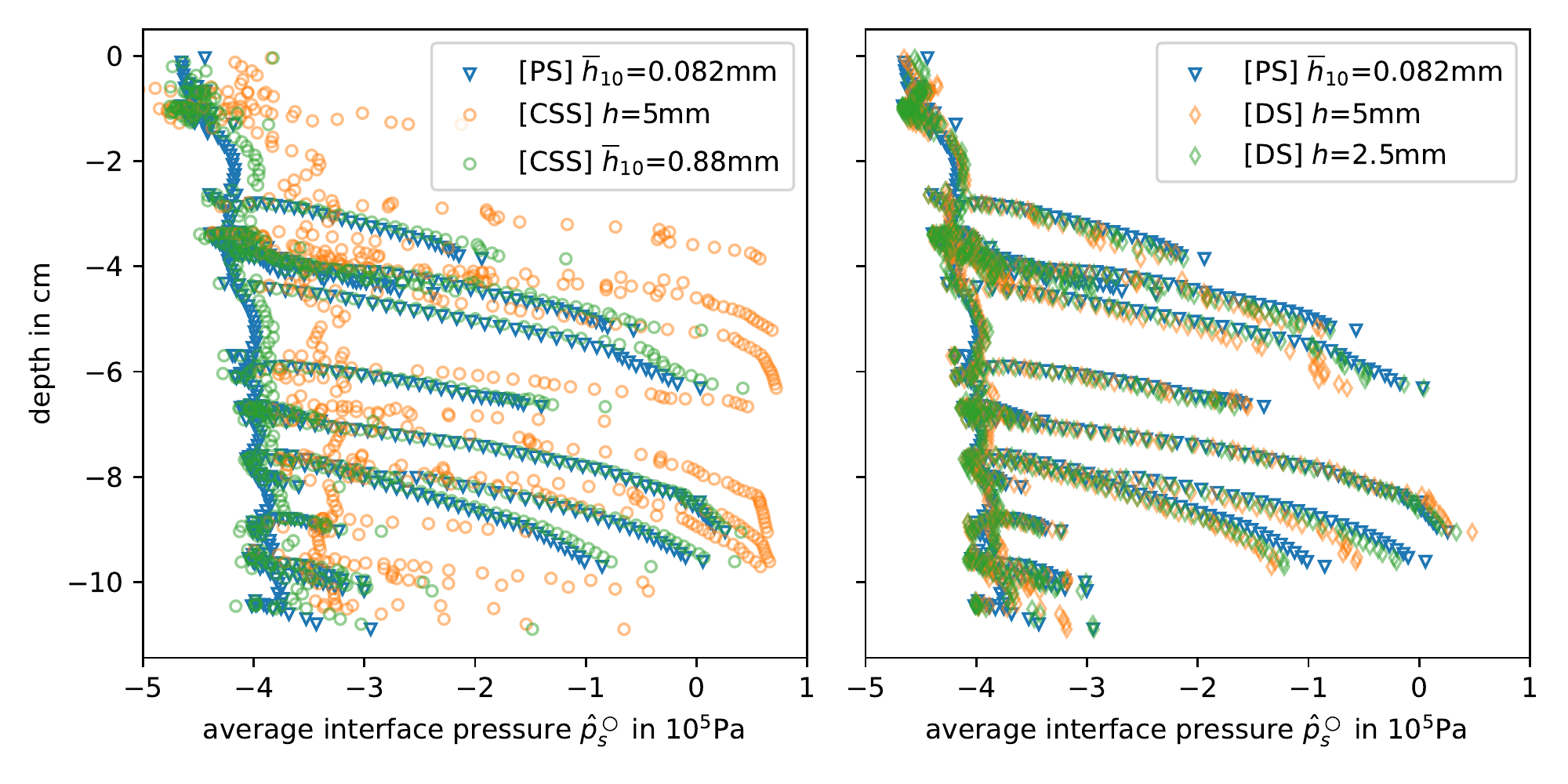}
	\caption{\textbf{Average soil pressure on the interface.} $\hat{p}_s^\bigcirc$
	for different methods and $p_{r,c} = \SI{-5e5}{\pascal}$.
	[DS] distributed source method (this paper),
	[CSS] cylinder surface method, cf.~\citep{koeppl2018,Koch2018a},
	[PS] projection method, resolved root-soil interface, cf.~\citep{Koch2021a}.
	The symbol $\overline{h}_{10}$ denotes the average diameter of the smallest \SI{10}{\percent} of the cells.
	The DS method uses a uniform mesh with cell diameter $h \equiv \overline{h}_{10}$.
	The values are computed per root segment.
	Meshes for CSS are $8$k, $1.9$M cells, for DS $8$k, $61$k cells, and for PS $12$M cells, respectively.}
	\label{fig:pbarsoil}
\end{figure}
\Cref{fig:pbarsoil} shows the reconstructed interface soil pressures $\hat{p}_s^\bigcirc$
for all root segments in the mesh over the domain depth for the case $p_{r,c} = \SI{-5e5}{\pascal}$.
For PS, $\hat{p}_s^\bigcirc$ is computed as a numerical integral over all
coupling surface facets; for CSS as numerical integral over the cylinder surface; for DS
it is the reconstructed value using the proposed reconstruction algorithm \cref{eq:nleqifpressure-mult}.
The PS solution is considered a reference, although the results in \cref{fig:transpiration} suggest
that the result is not fully converged, the results can be considered quite accurate with
less than $\SI{3}{\percent}$ difference in the transpiration rate to the DS method and the CSS method with
strong local grid refinement. It can be seen that for the CSS method and the coarse grid,
the interface soil pressure is overestimated explaining the large overestimation
of the transpiration rate. With local grid refinement there is a much closer match with the PS
solution. However, we note that in particular for the larger roots (low $\hat{p}_s^\bigcirc$),
difference between PS and CSS are still clearly visible. On the other hand,
the DS closely matches the PS solution even for a coarse grid discretization.
The largest difference is observed towards the root tips (highest $\hat{p}_s^\bigcirc$).
The difference is improved by grid refinement. Note that at $h=\SI{2.5}{\mm}$ and
given that $\varrho_i = 3R_i$ the kernel support around most smaller root segments
is hardly resolved by one mesh cell, cf. \cref{fig:root-conductivities-geometry}.
Therefore, this close match between the novel DS method and
the PS method using a fully-resolved root-soil interface and more than a thousand times more mesh cells
is remarkable. This shows that the root water uptake problem, in particular in drier soils
is clearly dominated by large and very localized gradients at the root-soil interface
that are difficult to approximate for standard numerical schemes.

Finally, we remark that there is a strong reduction in computational time
associated with the advantage of using a coarse grid
discretization and a structured cube grid for the three-dimensional domain.
While the DS method used \SI{1}{\s} (wall-clock time) per simulation for the $8k$ grid (\SI{5}{\min} for $4$M cells),
the interface resolving PS method used \SI{2.7}{\hour} for $11$M cells.
The DS method also required on average less Newton iterations. This can be attributed to the fact
that the nonlinearity in $D_b$ is shifted into the reconstruction algorithm, whereas the numerical solution
is regularized. For the projection method
the large interface gradients have to be approximated by the discrete solution.

\section{Summary and final remarks}
\label{sec:summary}

Mixed-dimensional methods are efficient methods for solving coupled
mixed-dimensional PDEs arising from flow and transport processes in systems with tubular network systems
embedded in a bulk domain. The bulk is represented by a three-dimensional mesh,
the tubes are given as a network of cylinder center-line segments, and the meshes are typically non-matching.
If the diffusion coefficient in the bulk domain depends on the unknown concentration a coupled nonlinear diffusion equation has to be solved.
Its solution may exhibit large local gradients at the tube-bulk interface.
This is for example the case when modeling water transport in soils with embedded
root systems. Roots take up water from the soil and transport it upwards toward the atmosphere.
In particular in dry soils, water uptake causes a strong local drop in the hydraulic conductivity
leading to large pressure gradients around root segments.

We introduced an efficient numerical method for the solution of such nonlinear mixed-dimensional PDEs.
The method is based on source distribution in a finite neighborhood region of the network,
in combination with a nonlinear reconstruction scheme for interface unknowns. The method
is based on several approximations we have analyzed. We estimated the errors associated with the approximations
and showed in series of numerical verification tests that
these errors remain small in practical applications.
We use the new method to simulate root water uptake using a realistic root network.
In comparison with existing methods we showed that the novel method outperforms
other methods in both accuracy and efficiency. While the numerical results clearly show
that the method accurately solves stationary problems, time-dependent problems are yet
to be investigated in future work. However, preliminary results with root water uptake
and slowly varying conditions (e.g. diurnal cycles) suggest that the method remains accurate.

Due to the possibility to use coarse computational grids
in comparison with the tube diameter of the embedded networks, the presented method
allows to perform simulations with large networks with reasonable 3D grid resolutions.
Coarse grid discretizations ($h \gg R$) in mixed-dimensional methods effectively lead to
a distribution of any exchange source term $q$ of an embedded tube into a neighborhood of diameter $h$.
However, since local variations of a bulk unknown $u_b$ cannot be resolved, the approximation of a source
term depending on $u_b$ evaluated on the tube-bulk interface suffers from significant errors.
By deliberately introducing a distribution kernel for the source term,
$u_b$ is locally modified and deviates from the real solution. However, by construction, the behavior of
$u_b$ in the vicinity of the tube-bulk interface is better understood and it is possible to
develop an interface reconstruction scheme, see \cref{eq:nleqifpressure} and \cref{sec:reconstruction}.
We have shown that the reconstruction significantly reduces the discretization error.
Another approach is the Peaceman well model known in reservoir engineering \citep{Peaceman1978,Peaceman1983,Koch2019bwell}.
Peaceman devises a reconstruction method eliminating discretization errors for one specific discretization scheme,
and with some assumptions on the structure of the mesh and the orientation of the well tube. Methods based
on the discrete formulation have also been explored for root water uptake simulations~\citep{Schroeder2008localsoil,Beudez2013,Mai2019}.
In contrast, the distributed source approach used in this work is formulated
in the continuous setting which allows the analysis of the problem from a different perspective,
and is applicable for any discretization scheme. Furthermore, the model remains valid in the
case that the discretization length is smaller than the tube radius, which may readily occur if the
tube radii vary significantly in the network (e.g large root systems).
The distribution kernel effectively relaxes the strong discretization length restriction
on the tube radius present in methods that require a direct numerical approximation of the interface unknown~\citep{Koch2019a}.
Our observations extend the results of \citep{Koch2019a} to the nonlinear case and indicate that as soon as
the discretization length is in the range of the kernel radius ($h \approx \varrho$) the error in the source
term is already reasonably small for many application scenarios. In effect, this property
allows to improve the solution accuracy for simulations where finer grid discretization are not feasible, e.g.
because of high computational costs.

\section*{Acknowledgements}

We want to thank Kent-André Mardal and Rainer Helmig for constructive comments on the initial manuscript.
This work was financially supported by the German Research Foundation (DGF),
within the Collaborative Research Center on
Interface-Driven Multi-Field Processes in Porous Media (SFB 1313, Project Number 327154368).
T. Koch also acknowledges funding
from the European Union's Horizon 2020 research and innovation programme
under the Marie Skłodowska-Curie grant agreement No 801133.

\section*{CRediT author statement}
\textbf{T. Koch:} Conceptualization, Methodology, Software, Validation, Investigation, Data Curation, Writing -- Original Draft, Visualization, Supervision, Project administration
\textbf{H. Wu:} Software, Validation, Investigation, Data Curation, Writing -- Review \& Editing, Visualization
\textbf{M. Schneider:} Conceptualization, Methodology, Formal analysis, Writing -- Original Draft, Supervision, Project administration

\begin{appendices}

\section{Exponential diffusion coefficient function}
\label{sec:app:exp}

For numerical verification tests, we choose the following exponential function for $D_b(u_b)$
as an example for a strongly nonlinear function (depending on the rate parameter $k$),
\begin{equation}
\begin{split}
	D_b(u_b) &= \max( D_0 \exp\{ k (u_b - 1) \}, D_\text{min} ) \\
	&=
	\begin{cases}
		D_\text{min} & u_b \leq u_c \\
		D_0 \exp\{ k (u_b - 1) \} & u_b > u_c
	\end{cases},
\end{split}
\end{equation}
where $D_0$ is a constant diffusion coefficient, $u_c = 1 + \frac{1}{k}\ln\{ \frac{D_\text{min}}{D_0} \}$ such that $D_b$ is continuous,
and $D_\text{min} = D_0 \epsilon$ (where $\epsilon > 0$ is a small constant) is a minimal diffusion coefficient,
for the purpose of rendering the Kirchhoff transformation, $T(u_b)$, and its inverse, $T^{-1}(\psi)$, well-defined on all of $\mathbb{R}$ (see \cref{sec:kirchhoff}).



We obtain analytical expressions for $T(u_b)$ and $T^{-1}(\psi)$ by splitting the integral range
depending on the sign of $u_c$ and depending which one of $u_b$ and $u_c$ is larger.
The Kirchhoff transformation, as defined in \cref{eq:kirchhoff},
is given for $u_c \leq 0$ by
\begin{equation}
	T(u_b) = \begin{cases}
		 D_\text{min}(u_b-u_c) +  \frac{D_0}{k} \left(D_{b,r}(u_c) -  D_{b,r}(0) \right)  & u_b \leq u_c \\
		 \frac{D_0}{k} \left( D_{b,r}(u_b) - D_{b,r}(0)\right)  & u_b > u_c, \\
	\end{cases}
\end{equation}
and for $u_c > 0$ by
\begin{equation}
	T(u_b) = \begin{cases}
		 D_\text{min}u_b  & u_b \leq u_c \\
		\frac{D_0}{k} \left( D_{b,r}(u_b) - D_{b,r}(u_c) \right) +  D_0 D_{b,r}(u_c) u_c & u_b > u_c, \\
	\end{cases}
\end{equation}
where $D_{b,r}(u_b) = \exp\{ k (u_b - 1) \}$. The inverse transformation, for the case that $u_c \leq 0$, is given by
\begin{equation}
	T^{-1}(\psi) = \begin{cases}
		\frac{1}{D_\text{min}}\left(\psi  - \frac{D_0}{k} \left(D_{b,r}(u_c) -  D_{b,r}(0) \right) \right) + u_c & \psi \leq T(u_c) \\
		1 + \frac{1}{k}\ln\left\{\frac{k}{D_0} \psi  + D_{b,r}(0)   \right\} & \psi > T(u_c), \\
	\end{cases}
\end{equation}
and for the case that $u_c > 0$, by
\begin{equation}
	T^{-1}(\psi) = \begin{cases}
		\frac{1}{D_\text{min}}\psi  & \psi \leq T(u_c) \\
		1 + \frac{1}{k}\ln\left\{\frac{k}{D_0} (\psi - D_0 D_{b,r}(u_c)u_c)  + D_{b,r}(u_c)   \right\} & \psi > T(u_c). \\
	\end{cases}
\end{equation}
The functions $D_b(u_b)$, $T(u_b)$, and $T^{-1}(\psi)$ are plotted for $D_0 = 1$ and $\epsilon = \SI{1e-6}{}$ in \cref{fig:Dbr}.

\section{Error estimate for arbitrarily-oriented tubes in 3D}
\label{sec:app:estimate_3d}

In the following we want to derive an estimate for $|\avg{f_j}{\per_i} - f_j(\vec{x}_i)|$.
Using the multivariate Taylor expansion for $f_j$ for any  $\vec{x} \in B_{R_i}(\vec{x}_i)$ yields
\begin{equation}
f_j(\vec{x}) = f_j(\vec{x}_i) + \nabla f_j(\vec{x}_i) \cdot (\vec{x} - \vec{x}_i) + \mathcal{R}_2(\vec{x} - \vec{x}_i),
\label{eq:TaylorEx}
\end{equation}
with the second-order residual $|\mathcal{R}_2(\vec{x} - \vec{x}_i)| \leq \frac{C}{2} \lVert \vec{x} - \vec{x}_i\rVert^2_2 $
and the constant
\begin{equation}
C:=\sup\limits_{|\alpha| = 2}\left( \sup\limits_{\vec{x} \in B_{R_i}(\vec{x}_i)} |\partial^\alpha f_j(\vec{x})| \right).
\end{equation}
Applying the averaging operator to \cref{eq:TaylorEx} results in
\begin{equation}
|\avg{f_j}{\per_i} - f_j(\vec{x}_i)| \leq \frac{C}{2} R_i^2.
\end{equation}

We then assume that the contributions from neighboring tubes
contributions from all neighboring tubes $j$ can be formulated in terms of line sources
and that the tubes are non-overlapping. The influence of segment sources decays faster with
the distance than for the line source, so we consider this a conservative assumption.
The corresponding functions $f_j$ are given by
\begin{align}
\label{eq:fjSolA}
f_j(\vec{x}) &= \begin{cases} \frac{1}{2\pi}\left[ \frac{|| \vec{x} - \mathcal{E}_j^\perp(\vec{x}) ||_2^2}{2\varrho_j^2} + \ln\left(\frac{\varrho_j}{R_j}\right) - \frac{1}{2} \right] & ||\vec{x} - \mathcal{E}_j^\perp(\vec{x})||_2 \leq \varrho_j, \\
\frac{1}{2\pi }\left[ \ln\left(\frac{|| \vec{x} - \mathcal{E}_j^\perp(\vec{x}) ||_2}{R_j}\right) \right] & \text{else,}
\end{cases}
\end{align}
where $\mathcal{E}_j^\perp$ orthogonally projects $\vec{x}$ onto the centerline of tube $j$.

For such functions (assuming non-overlapping kernel support regions),
the constant $C$ can explicitly be estimated by calculating the second-order derivatives.
We note that gradients of the assumed $f_j$ are aligned with the radial direction
of a cylinder coordinate system $(r,\theta,s)$ implied by centerline $j$. Therefore, it is enough to analyse radial derivatives of $f_j$:
\begin{equation}
\frac{\partial f_j}{\partial r} = \frac{1}{2\pi r}, \quad \frac{\partial^2 f_j}{\partial r^2} = -\frac{1}{2\pi r^2}.
\end{equation}
With the definition of $C$, this results in the estimate
\begin{equation}
\begin{split}
|\avg{f_j}{\per_i} - f_j(\vec{x}_i)| &\leq  \frac{R_i^2}{4\pi\inf\limits_{\vec{x} \in B_{R_i}(\vec{x}_i)} \lVert \vec{x} - \mathcal{E}_j^\perp(\vec{x}) \rVert_2^2 } \\
                                     &= \frac{R_i^2}{4\pi (\lVert \vec{x}_i - \mathcal{E}_j^\perp(\vec{x}_i) \rVert_2 - R_i)^2 },
\end{split}
\end{equation}
where for the last equality we have used the assumption that $\mathcal{E}_j^\perp(\vec{x}_i) \not \in B_{R_i}(\vec{x}_i)$.




\section{Analytical solution for three tubes ($\varrho_i = 2R_i$)}
\label{sec:app:tworho}

\begin{figure}[!htb]
	\includegraphics[width=1.0\textwidth,trim=60 50 50 50,clip]{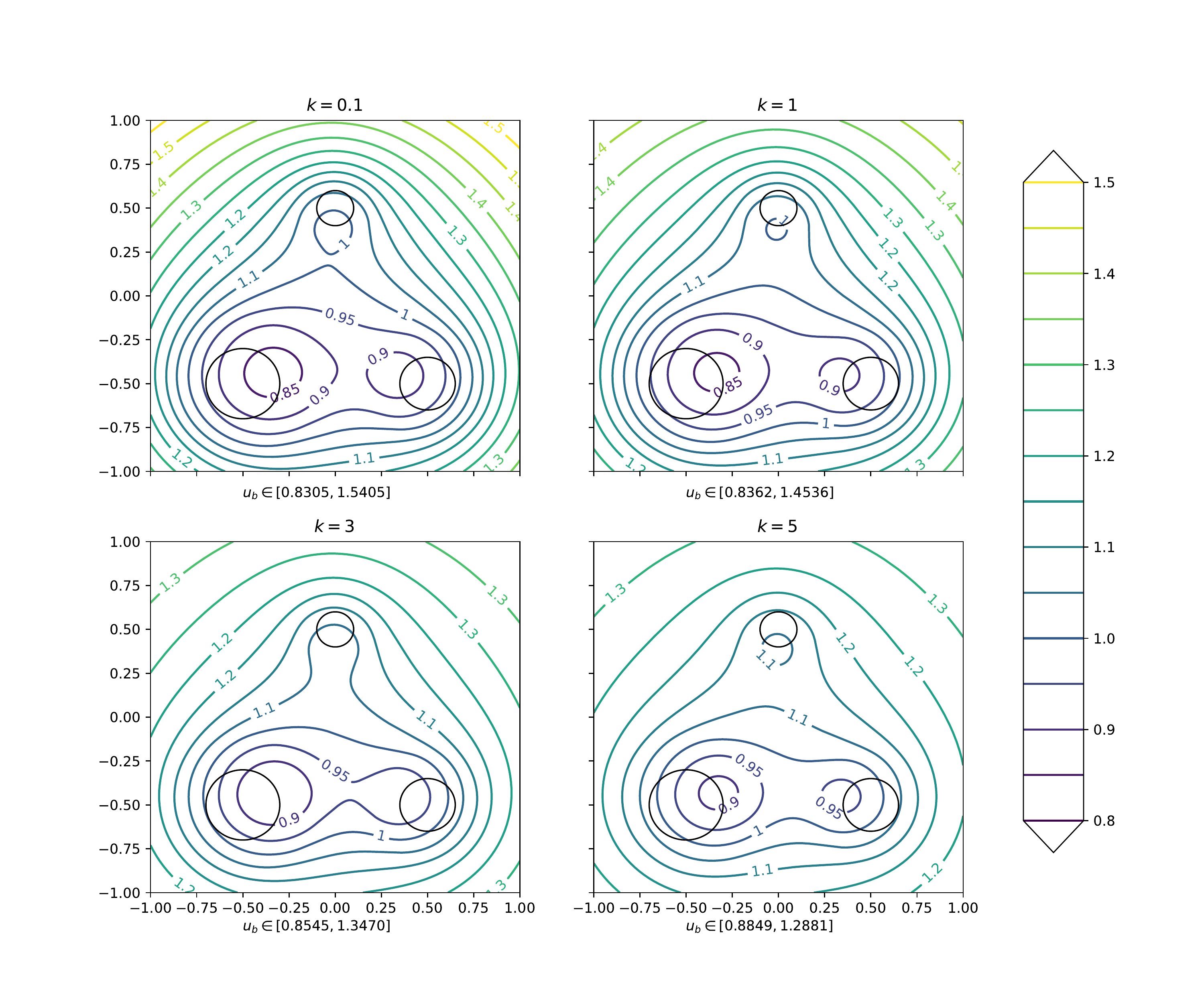}
	\caption{Contour plot of the analytical solution $U_b$
		for three parallel tubes (cross-sectional perimeter shown by black circles)
		with radii $R = (0.2, 0.15, 0.1)$, $\varrho_i = 2 R_i$,
		and for various exponential diffusion coefficient functions (\cref{fig:Dbr}, $D_0 = 0.5$). The minimum and maximum values of $u_b$ are given under each plot.
	}
	\label{fig:roots_2R}
\end{figure}

In \cref{fig:roots3}, we show an analytical solution for three point sources plotted
for $\varrho_i = R_i$ to show the variations of $\hat{u}_b$ over the tube-bulk interface that determine
the approximation error of the interface reconstruction scheme. However, the numerical tests
are run for the analytical solution with $\varrho_i = 2R_i$ which suffer from the same approximation error
of the interface reconstruction scheme but $u_b$ is much smoother due to the distribution kernel.
To show how much influence this actually has on very nonlinear $D_b$ we here show in \cref{fig:roots_2R} the solution for
$\varrho_i = 2R_i$. As noted in the introduction, the distribution kernel
turns a difficult to approximate function $\hat{u}_b$ into a much smoother $u_b$ better suited for numerical approximation.
However, using the developed reconstruction algorithm, the correct source term $q$ can be reconstructed
from the regularized solution.
\end{appendices}

\bibliography{kernel}
\bibliographystyle{elsarticle-num}

\end{document}